\documentclass[prd,twocolumn,nofootinbib]{revtex4}

\usepackage{amsmath,amssymb}
\usepackage{graphicx}
\usepackage{pst-plot}
\usepackage{mathrsfs}
\usepackage{mathtools}
\usepackage{pst-func}
\usepackage{rotating}

\newtheorem{theorem}{Theorem}

\newtheorem{hypothesis}{Hypothesis}

\def\dbl{\hbox{${1\hskip -2.4pt{\rm l}}$}}

\begin{document}

\title{Symmetric derivation of singlet correlations in a quaternionic 3-sphere model}

\author{Joy Christian}

\email{jjc@bu.edu}

\affiliation{Einstein Centre for Local-Realistic Physics, Oxford OX2 6LB, United Kingdom}

\begin{abstract}
Using the powerful language of geometric algebra, we present an observationally symmetric derivation of the strong correlations predicted by the entangled singlet state in a deterministic and locally causal model, usually also referred to as a local-realistic model, in which the physical space is assumed to be a quaternionic 3-sphere, or $S^3$, available as the spatial part of a solution of Einstein's field equations of general relativity, and compare it in quantitative detail with Bell's local-realistic model for the singlet correlations set within a flat Euclidean space ${\mathrm{I\!R}^3}$. Since the quantitatively detailed expressions of relative-angle-dependent probabilities of observing measurement outcomes for Bell's local model do not seem to have been fully articulated before, our novel analysis exploiting the non-commutative properties of quaternions, in addition to allowing the comparison with the quaternionic 3-sphere model, may also provide useful comparisons for other less compelling local-realistic models, such as those relying on retrocausality or superdeterminism. Apart from the conservation of zero spin angular momentum, the key attribute underlying the strong singlet correlations within $S^3$ in comparison with Bell's local model turns out to be the spinorial sign changes intrinsic to quaternions that constitute the 3-sphere. In addition, we also discuss anew a macroscopic experiment that can, in principle, test our 3-sphere hypothesis.
\end{abstract}

\maketitle

\baselineskip 12.9pt

\parskip 4pt

\section{Introduction}

It is well known that a cosmological solution of Einstein's field equations of general relativity known as the Friedmann-Lema\^itre-Robertson-Walker spacetime admits three possible geometries for the three-dimensional physical space, isomorphic to ${\mathrm{I\!R}^3}$, $H^3$, or $S^3$, where ${\mathrm{I\!R}^3}$ represents an open Euclidean space, $H^3$ represents an open hyperboloid of negative curvature, and $S^3$ represents a closed 3-sphere of constant positive curvature \cite{d'Inverno}. Among these possible geometries, only $S^3$ represents a closed universe with compact geometry. Moreover, observationally, the cosmic microwave background spectra mapped by the space observatory {\it Planck} now prefers a positive curvature, or $S^3$, at more than 99\% confidence level \cite{closed,Handley}.

On the other hand, Bell's argument (or Bell's theorem \cite{Bell-1964}) against possible local-realistic models for quantum correlations is set within a flat and immutable spacetime. Indeed, in Chapter 7 of his book \cite{Speakable}, while exploring possible strategies to overcome his argument, Bell wonders: ``The space time structure has been taken as given here. How then about gravitation?'' Considering this shortcoming of Bell's argument and other physical considerations in the foundations of quantum mechanics, together with the universal validity of Einstein's field equations of general relativity at all macroscopic scales including terrestrial and other non-cosmological scales, in a series of works since 2007 \cite{Disproof,IJTP,RSOS,IEEE-1,IEEE-2,local,RSOS-Reply} we have proposed the following experimentally testable hypothesis:
\begin{hypothesis} \label{hyp1}
The strong quantum correlations observed in Nature can be understood as manifestly local, realistic, and deterministic correlations if we model the three-dimensional physical space as a closed and compact quaternionic 3-sphere $S^3$ using geometric (or Clifford) algebra \cite{Clifford} rather than as a flat and open space ${\mathrm{I\!R}^3}$ using ``vector algebra.''
\end{hypothesis}
Here, the correlations are said to be strong if the absolute value of the Bell-CHSH combination \cite{Bell-1964,Speakable} of expectation values 
\begin{equation}
{\cal E}({\bf a},\,{\bf b})+{\cal E}({\bf a},\,{\bf b'})+{\cal E}({\bf a'},\,{\bf b})-{\cal E}({\bf a'},\,{\bf b'}) \label{combi}
\end{equation}
calculated for them exceeds the bound of $2$, where the functions ${\cal E}({\bf a},\,{\bf b})$ relevant for us here are defined in equation (\ref{for-1}) below. Moreover, the geometry of physical space we are concerned with here is its {\it intrinsic} geometry at the terrestrial, solar, or galactic scales \cite{IEEE-1}. Extrinsic geometries, cosmological scales, and local gravitational interactions will not be relevant to our analysis. In other words, it is not the strength or weakness of gravitational interactions that will be relevant for us here but the qualitative algebraic differences between the geometries of $S^3$ and $\mathrm{I\!R}^3$, which are admissible spatial parts of the solution of Einstein's field equations of general relativity we alluded to above. On the other hand, geometrically the tangent space ${T_p S^3}$ at each point $p$ of $S^3$ is isomorphic to ${{\mathrm{I\!R}}^3}$, analogous to how the tangent space at each point of the surface $S^2$ of a ball is isomorphic to the plane ${{\mathrm{I\!R}}^2}$. Moreover, the tangent bundle $TS^3$ of ${S^3}$ is trivial:
\begin{equation}
TS^3:=\!\bigcup_{p\in S^3}\{p\}\times T_p S^3 = S^3\times{\mathrm{I\!R}}^3.\label{cfeq3}
\end{equation}
Therefore, locally the space $S^3$ is indistinguishable from ${{\mathrm{I\!R}}^3}$. As a result, the local experiences of experimenters within $S^3$ will be no different from those of their counterparts within ${{\mathrm{I\!R}}^3}$, the flat geometry of which is usually taken for granted in all investigations and analyses concerning Bell's theorem. Our hypothesis is thus similar to ancient suspicions that Earth is spherical despite the local appearance suggesting that it is flat.

The above Hypothesis~\ref{hyp1} has far-reaching consequences, some of which we have explored previously in \cite{Disproof,IJTP,RSOS,IEEE-1,IEEE-2,local,RSOS-Reply}. Our goal here is to explicitly derive one of these consequences using the powerful language of geometric algebra, namely the strong correlations predicted by the entangled singlet state within the local-realistic framework of Bell \cite{Bell-1964}, and compare it, in quantitative details, with Bell's local model\footnote{In this paper we are only concerned with the probabilities of observing measurement outcomes for the explicit local-realistic model presented by Bell in Section~3 of \cite{Bell-1964} and not directly with the so-called theorem he has presented in the same paper. The latter we have criticized elsewhere, such as in Refs.~\cite{RSOS,Begs,Oversight}. In Sections~II and III of \cite{Begs}, a succinct review of what is meant by local realism and hidden variable theories can also be found.\label{foot-1}} for the singlet correlations set within a flat Euclidean space ${\mathrm{I\!R}^3}$. This requires a different set of hidden variables from the one employed in \cite{Disproof,IJTP,RSOS,IEEE-1,IEEE-2,local,RSOS-Reply} (cf. footnote~\ref{foot-2} below). Since, as far as we are able to ascertain, the probabilities of observing measurement outcomes for Bell's local model have never been fully worked out before, our novel analysis exploiting the non-commutative properties of quaternions may also provide useful comparisons for other less compelling local-realistic models, such as those relying on retrocausality and superdeterminism. In addition, we will also discuss anew a macroscopic experiment proposed in \cite{IJTP} that can, in principle, test the above hypothesis.

To that end, let us begin by defining a quaternionic 3-sphere in the language of Geometric Algebra \cite{Clifford} whose tangent spaces are locally ${\mathrm{I\!R}^3}$:
\begin{equation}
S^3:=\left\{\,{\bf q}(\beta,\,{\mathbf r})=\varrho_r\exp\left[{\mathbf J}({\mathbf r})\,\frac{\beta}{2}\right]
\Bigg|\;\left|\left|\,{\bf q}\left(\beta,\,{\mathbf r}\right)\,\right|\right|=\varrho_r\right\}, \label{nonsin}
\end{equation}
where
\begin{equation}
\exp\left[{\mathbf J}({\mathbf r})\,\frac{\beta}{2}\right]=\cos\left(\frac{\beta}{2}\right)+{\mathbf J}({\mathbf r})\,\sin\left(\frac{\beta}{2}\right),
\end{equation}
${{\mathbf J}({\mathbf r})=I_3{\mathbf r}}$ is a unit bivector (or pure quaternion) rotating about a unit axis vector ${{\bf r}\in{\mathrm{I\!R}}^3}$ with rotation angle ${0\leq\beta < 4\pi}$, $\varrho_r$ is the radius of the 3-sphere, and $I_3={\mathbf e}_x\wedge{\mathbf e}_y\wedge{\mathbf e}_z={\mathbf e}_x{\mathbf e}_y{\mathbf e}_z$ is the standard trivector in the Clifford algebra $\text{Cl}_{3,0}$ of orthogonal directions in ${\mathrm{I\!R}^3}$ \cite{Clifford}. The basis bivectors of ${{\mathbf J}({\mathbf r})}$ generate the even subalgebra of $\text{Cl}_{3,0}$:
\begin{equation}
\mathbf{J}_{j}\,\mathbf{J}_{k} =-\,\delta_{jk}\,-\sum_l\epsilon_{jkl}\,\mathbf{J}_{l}\,, \label{bi-1-m}
\end{equation}
where ${\delta_{jk}}$ is the Kronecker delta and ${\epsilon_{jkl}}$ is the Levi-Civita symbol.
Note that this bivector subalgebra is isomorphic to the familiar algebra of Pauli matrices ${\boldsymbol{\sigma}_j}$
(${j=x,y,z}$) specified by
\begin{equation}
\boldsymbol{\sigma}_j\boldsymbol{\sigma}_k=\,\delta_{jk}\,{\dbl}\,+\,i\,\sum_l\epsilon_{jkl}\,\boldsymbol{\sigma}_l\,, \label{paulialgebra}
\end{equation}
where $\dbl$ is $2\times2$ identity matrix and ${i\equiv\sqrt{-1}}$ is a unit imaginary. The isomorphism between the algebras (\ref{bi-1-m}) and (\ref{paulialgebra}) can be easily established by the transformations ${\mathbf{J}_k\Longleftrightarrow{i\boldsymbol{\sigma}_k}}$ and $\dbl\Longleftrightarrow1$.

It is easy to verify that quaternions ${\bf q}(\beta,\,{\mathbf r})$ appearing in (\ref{nonsin}) respect the rotational symmetries exhibited by spinors:
\begin{equation}
{\bf q}(\beta+2\kappa\pi,\,{\bf r})=(-1)^{\kappa}\,{\bf q}(\beta,\,{\bf  r})\;\;\,\text{for}\;\,\kappa=0,1,2,3,\dots \label{signchanges}
\end{equation}
Thus, we can use ${{\bf q}(\beta,\,{\bf r})}$ to represent states of a physical system that returns to itself only after even multiples of ${2\pi}$ rotations. Given two unit vectors ${\bf x}$ and ${\bf y}$ and a rotation axis ${\bf r}$, each of the quaternions ${\bf q}(\eta_{{\bf x}{\bf y}},\,{\bf r})$ in a unit ${S^3}$ can be factorized into a product of corresponding bivectors ${{\bf J}({\bf x})}$ and ${{\bf J}({\bf y})}$ [which can be expanded using the basis bivectors in (\ref{bi-1-m})] as follows:
\begin{align}
{\bf q}(\eta_{{\bf x}{\bf y}},\,{\bf r})&=-\,{\bf J}({\bf x})\,{\bf J}({\bf y})=-\,(I_3{\bf x})\,(I_3{\bf y}) \label{20} \\
&=-(I_3)^2{\bf x}\,{\bf y}={\bf x}\,{\bf y}={\bf x}\cdot{\bf y}\,+\,{\bf x}\wedge{\bf y} \\
&=\cos(\eta_{{\bf x}{\bf y}})+{\bf J}({\bf r})\,\sin(\eta_{{\bf x}{\bf y}}), \label{22}
\end{align}
which are now evidently unit quaternions, $||{\bf q}(\eta_{{\bf x}{\bf y}},\,{\bf r})||=1$, if ${\eta_{{\bf x}{\bf y}}}$ is the angle between ${\bf x}$ and ${\bf y}$, ${{\bf x}\,{\bf y}}$ is the geometric product between ${\bf x}$ and ${\bf y}$, ${{\bf x}\wedge{\bf y}}$ is the wedge product between ${\bf x}$ and ${\bf y}$, and ${{\bf J}({\bf r})}$ is identified with ${\frac{{\bf x}\wedge{\bf y}}{||{\bf x}\wedge{\bf y}||}}$. Comparing ${\bf q}(\eta_{{\bf x}{\bf y}},\,{\bf r})$ in (\ref{20}) with ${{\bf q}(\beta,\,{\bf r})}$ defined in (\ref{nonsin}) for $\varrho_r=1$, we see that the rotation angle ${\beta}$ of ${{\bf q}(\beta,\,{\bf r})}$ is twice the angle ${\eta_{{\bf x}{\bf y}}}$ between the vectors ${\bf x}$ and ${\bf y}$ in any factorization such as that in (\ref{20}):
\begin{equation}
\beta=2\,\eta_{{\bf x}{\bf y}}.
\end{equation}
As a result, the spinorial sign changes exhibited by quaternions as shown in (\ref{signchanges}) can be expressed also using $\eta_{{\bf x}{\bf y}}$ as
\begin{equation}
{\bf q}(\eta_{{\bf x}{\bf y}}+\kappa\pi,\,{\bf r})=(-1)^{\kappa}\,{\bf q}(\eta_{{\bf x}{\bf y}},\,{\bf r})\,\;\;\text{for}\;\,\kappa=0,1,2,3,\dots \label{spinorial}
\end{equation}
As we shall demonstrate below, relation (\ref{signchanges}) exhibits the key property that induces the singlet correlations within $S^3$. 

\section{Locally causal, realistic, and deterministic measurement functions}

\begin{figure*}[t]
\hrule
\scalebox{0.9}{
\begin{pspicture}(0.0,-2.4)(5.4,2.4)

\psline[linewidth=0.1mm,dotsize=3pt 4]{*-}(-2.51,0)(-2.5,0)

\psline[linewidth=0.1mm,dotsize=3pt 4]{*-}(7.2,0)(7.15,0)

\psline[linewidth=0.4mm,arrowinset=0.3,arrowsize=3pt 3,arrowlength=2]{->}(-2.5,0)(-3,1)

\psline[linewidth=0.4mm,arrowinset=0.3,arrowsize=3pt 3,arrowlength=2]{->}(-2.5,0)(-3,-1)

\psline[linewidth=0.4mm,arrowinset=0.3,arrowsize=3pt 3,arrowlength=2]{->}(7.2,0)(8.3,0.5)

\psline[linewidth=0.4mm,arrowinset=0.3,arrowsize=3pt 3,arrowlength=2]{->}(7.2,0)(7.4,1.3)

\psline[linewidth=0.4mm,arrowinset=0.3,arrowsize=2pt 3,arrowlength=2]{->}(4.2,0)(4.2,1.1)

\psline[linewidth=0.4mm,arrowinset=0.3,arrowsize=2pt 3,arrowlength=2]{->}(0.5,0)(0.5,1.1)

\pscurve[linewidth=0.2mm,arrowinset=0.2,arrowsize=2pt 2,arrowlength=2]{->}(4.0,0.63)(3.85,0.45)(4.6,0.5)(4.35,0.65)

\put(4.1,1.25){{\large ${{\bf s}_2}$}}

\pscurve[linewidth=0.2mm,arrowinset=0.2,arrowsize=2pt 2,arrowlength=2]{<-}(0.35,0.65)(0.1,0.47)(0.86,0.47)(0.75,0.65)

\put(0.4,1.25){{\large ${{\bf s}_1}$}}

\put(-2.4,+0.45){{\large ${\bf 1}$}}

\put(6.8,+0.45){{\large ${\bf 2}$}}

\put(-3.35,1.35){{\large ${\bf a}$}}

\put(-3.5,-1.7){{\large ${\bf a'}$}}

\put(8.5,0.52){{\large ${\bf b}$}}

\put(7.3,1.5){{\large ${\bf b'}$}}

\put(1.8,-0.65){\large source}

\put(0.99,-1.2){\large ${\pi^0\longrightarrow\,e^{-}+\,e^{+}\,}$}

\put(1.11,0.5){\large total spin = 0}

\psline[linewidth=0.3mm,linestyle=dashed](-2.47,0)(2.1,0)

\psline[linewidth=0.4mm,arrowinset=0.3,arrowsize=3pt 3,arrowlength=2]{->}(-0.3,0)(-0.4,0)

\psline[linewidth=0.3mm,linestyle=dashed](2.6,0)(7.2,0)

\psline[linewidth=0.4mm,arrowinset=0.3,arrowsize=3pt 3,arrowlength=2]{->}(5.0,0)(5.1,0)

\psline[linewidth=0.1mm,dotsize=5pt 4]{*-}(2.35,0)(2.4,0)

\pscircle[linewidth=0.3mm,linestyle=dashed](7.2,0){1.3}

\psellipse[linewidth=0.2mm,linestyle=dashed](7.2,0)(1.28,0.3)

\pscircle[linewidth=0.3mm,linestyle=dashed](-2.51,0){1.3}

\psellipse[linewidth=0.2mm,linestyle=dashed](-2.51,0)(1.28,0.3)

\end{pspicture}}
\hrule
\caption{A spin-less neutral pion decays into an electron-positron pair. Measurements of spin components of each separated fermion are performed by experimenters Alice and Bob at remote stations ${\mathbf{1}}$ and ${\mathbf{2}}$, obtaining binary results $\mathscr{A}=\pm1$ and $\mathscr{B}=\pm1$ along directions ${\mathbf a}$ and ${\mathbf b}$. The conservation of spin angular momentum dictates that the total spin of the system remains zero during its free evolution after the fermions have separated and ceased to interact. Their initial rotation senses and directions are thus unaltered before they interact with the remote detectors to produce results. After \cite{IEEE-1}.}
\label{Fig-1}
\smallskip
\hrule
\end{figure*}
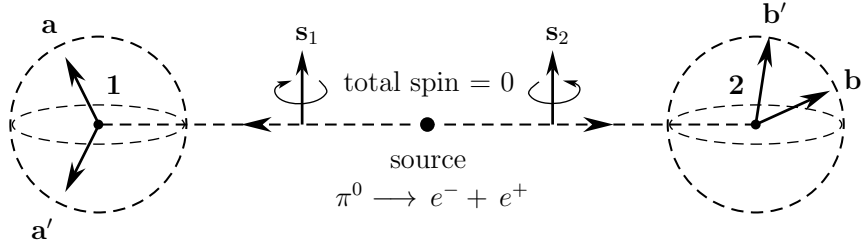

Now, for our purposes here, it is convenient to express the unit quaternions defined in (\ref{nonsin}) together with $\varrho_r=1$ as pairs of products of two bivectors such as ${\bf D}({\bf a})\,{\bf L}({\bf s}_1)$ and ${\bf D}({\bf b})\,{\bf L}({\bf s}_2)$, with unit bivectors ${\bf D}({\bf a})=I_3{\bf a}$ and ${\bf D}({\bf b})=I_3{\bf b}$ defined within $S^3$ about freely chosen unit detector directions ${\bf a}$ and ${\bf b}$ representing the detector orientations used by Alice and Bob, and unit bivectors ${\bf L}({\bf s}_1)=I_3{\bf s}_1$ and ${\bf L}({\bf s}_2)=I_3{\bf s}_2$ defined similarly about the unit spin directions ${\bf s}_1$ and ${\bf s}_2$ representing the spins detected along the detector orientations ${\bf a}$ and ${\bf b}$, as in a typical setup for Bell-test experiments \cite{Aspect} shown in Fig.~\ref{Fig-1}.

Recall that, geometrically \cite{IEEE-1,Clifford}, a bivector such as $I_3{\bf a}$ is an abstraction of a directed plane segment, with only a magnitude and a sense of rotation [counterclockwise ($+$) or clockwise ($-$)] as its properties. Neither the shape of the plane segment nor its rotation axis represented by the vector ${\bf a}$ is an intrinsic part of the bivector $I_3{\bf a}$. Thus, just as a unit vector ${\bf a}$ mathematically represents an orientation in $\mathrm{I\!R}^3$ or a point on $S^2$, a unit bivector $I_3{\bf a}$ represents an orientation in $\mathrm{I\!R}^4$ or a point on $S^3$. Therefore, a bivector such as ${\bf D}({\bf a})$ provides the correct representation of a detector orientation within $S^3$. On the other hand, as we noted above, the tangent space at each point of $S^3$ is isomorphic to $\mathrm{I\!R}^3$. Therefore, the local experiences of Alice and Bob in $S^3$ would be no different from those of their counterparts in $\mathrm{I\!R}^3$. In other words, Alice and Bob will not be aware of the fact that when they are choosing detector orientations ${\bf a}$ and ${\bf b}$ within $\mathrm{I\!R}^3$, they are in fact choosing bivectors ${\bf D}({\bf a})$ and ${\bf D}({\bf b})$ within $S^3$, just as our local experiences do not inform us that Earth is a sphere (for further discussion, see Appendix~\ref{QandA}).

Now, during the free evolution\footnote{ Here `free evolution' refers to the time evolution of a physical system in the absence of any external influences or interactions. For further explanation, see Appendix~\ref{QandA}.} of the two spins from the central source to detectors as shown in Fig.~\ref{Fig-1}, the conservation of the net zero angular momentum requires our representation of the total spin to respect the condition 
\begin{align}
-\,{\bf L}({\bf s}_1)\,+\,{\bf L}({\bf s}_2)=0 
\;&\Longleftrightarrow\;{\bf L}({\bf s}_1)={\bf L}({\bf s}_2) \notag \\
&\Longleftrightarrow\;{\bf s}_1=\,{\bf s}_2\,\equiv\,{\bf s}. \label{55}
\end{align}
Consequently, during free evolution, the spins do not change in either their senses or directions. Therefore we are justified in representing them by time-independent bivectors ${\bf L}({\bf s}_1)$ and ${\bf L}({\bf s}_2)$. Evidently, in the light of the product rule (\ref{bi-1-m}) for the unit bivectors, the above condition is equivalent to the condition
\begin{equation}
{\bf L}({\bf s}_1)\,{\bf L}({\bf s}_2)=\,I_3{\bf s}\,I_3{\bf s}=(I_3)^2\,{\bf s}\,{\bf s}=-1. \label{56}
\end{equation}
Next, we introduce the following two sign functions analogous to the ones introduced by Bell in his local model \cite{Bell-1964}:
\begin{equation}
\mu_1 =\text{sign}(\mathbf{a}\cdot\mathbf{s}^i_1)=\pm1
\;\;\;\text{and}\;\;\;
\mu_2 =\text{sign}(\mathbf{s}^i_2\cdot\mathbf{b})=\pm1, \label{79-nom}
\end{equation}
where the subscripts 1 and 2 on $\mu$ and ${\mathbf s}$ refer to the observation stations of Alice and Bob (cf. Fig.~\ref{Fig-1}), and the superscript $i$ on ${\mathbf s}$ indicates its initial direction at the source with respect to their chosen detector directions. The spin direction ${\mathbf s}={\mathbf s}_1={\mathbf s}_2$ acts as a hidden variable, just as in Bell's local model \cite{Bell-1964,Peres}. The function ${\mu_1=\mathrm{sign}({\mathbf a}\cdot{\mathbf s}^i_1)}$ can be understood as follows. If, initially ({\it i.e.}, before the detection process defined by the measurement functions to be specified below), the two unit vectors ${\mathbf a}$ and ${\mathbf{s}^i_1}$ happen to be pointing through the same hemisphere of $S^2\hookrightarrow\mathrm{I\!R}^3$ centered at the origin of ${\mathbf{s}^i_1}$, then $\mu_1=\mathrm{sign}({\mathbf a}\cdot{\mathbf s}^i_1)$ will be equal to $+1$, and if the two unit vectors ${\mathbf a}$ and ${\mathbf{s}^i_1}$ happen to be pointing through the opposing hemispheres of $S^2$ centered at the origin of ${\mathbf{s}^i_1}$, then $\mu_1=\mathrm{sign}({\mathbf a}\cdot{\mathbf s}^i_1)$ will be equal to $-1$, provided that ${{\mathbf a}\cdot{\mathbf s}^i_1\not=0}$. If ${{\mathbf a}\cdot{\mathbf s}^i_1}$ happens to be zero, then $\mu_1=\mathrm{sign}({\mathbf a}\cdot{\mathbf s}^i_1)$ will be assumed to be equal to the sign of the first nonzero component of ${\bf a}$ from the set ${\{a_x,\,a_y,\,a_z\}}$. And likewise for the function ${\mu_2=\mathrm{sign}({\mathbf s}^i_2\cdot{\mathbf b})}$.

With the above preliminaries in mind\footnote{It is worth noting that the hidden variable used in the 3-sphere model presented in \cite{RSOS,IEEE-1,IEEE-2} is quite different from the one used in the present model. In the model presented in \cite{RSOS,IEEE-1,IEEE-2} the orientation $\lambda=\pm1$ of $S^3$ is assumed to be a hidden variable. This implies that the bivector subalgebra (\ref{bi-1-m}) respected by the spin bivectors ${\mathbf L}({\mathbf s})$ and that respected by the detector bivectors ${\mathbf D}({\mathbf a})$ are {\it different}, but related by the orientation $\lambda$ so that
\begin{equation}
{\mathbf L}({\mathbf r})\,=\,\lambda\,{\mathbf D}({\mathbf r})\,\,\Longleftrightarrow\,\,{\mathbf D}({\mathbf r})\,=\,\lambda\,{\mathbf L}({\mathbf r})\,. \notag
\end{equation}
On the other hand, in the model presented here, the initial spin direction ${\mathbf s}^i$ is taken to be a hidden variable, just as it is in Bell's local model published in Section~3 of \cite{Bell-1964}. This choice of hidden variable is made here in order to facilitate a closer comparison of the $S^3$ model with Bell's local model in \cite{Bell-1964} based on $S^2\hookrightarrow\mathrm{I\!R}^3$. The difference between the two choices of hidden variables is at least algebraically quite significant. In the present model, the spin bivectors ${\mathbf L}({\mathbf s})$ and the detector bivectors ${\mathbf D}({\mathbf a})$ respect the {\it same} algebra defined by (\ref{bi-1-m}), so that, algebraically, ${\mathbf L}({\mathbf r})\equiv{\mathbf D}({\mathbf r})$.\label{foot-2}}, we can now state the central theorem, proved in several different ways in \cite{Disproof,IJTP,RSOS,IEEE-1,IEEE-2,local}.
\begin{theorem} \label{1}
The strong quantum mechanical correlations predicted by the entangled singlet state can be understood as classical, local, realistic, and deterministic correlations among the pairs of limiting scalar points ${{\mathscr A}({\bf a},\,{\bf s}^i_1)=\pm1}$ and ${{\mathscr B}({\bf b},\,{\bf s}^i_2)=\pm1}$ of a quaternionic 3-sphere, or $S^3$, assumed to be a model of the three-dimensional physical space.
\end{theorem}
Here the precise definition of what is meant by ``limiting scalar points ${{\mathscr A}({\bf a},\,{\bf s}^i_1)=\pm1}$'' is given in equations (\ref{79-nmn}) to (\ref{85-nom}) below. The proof of this theorem requires us to compute the correlations while preserving the geometrical properties of $S^3$:
\begin{equation}
{\cal E}({\mathbf a},\,{\mathbf b})\,=\int_{S^2}
{\mathscr A}({\mathbf a},\,{\bf s}^i_1)\,{\mathscr B}({\mathbf b},\,{\bf s}^i_2)\;p({\bf s}^i)\,d{\bf s}^i\,=-\cos(\,\eta_{{\mathbf a}{\mathbf b}}), \label{65a}
\end{equation}
where $S^2$ is the base manifold of $S^3$ \cite{IEEE-1,IEEE-2}, the function $p({\bf s}^i)$ specifies the normalized initial probability distribution of the spin direction ${\bf s}^i$ over $S^2$, and the functions ${\mathscr A}({\mathbf a},{\mathbf s}^i_1)$ and ${\mathscr B}({\mathbf b},{\mathbf s}^i_2)$ encode local physical interactions taking place during the detection processes at the two ends of the experiment, producing results observed by Alice and Bob. Since the non-uniform magnetic fields within detectors such as Stern–Gerlach devices are known to align spin directions of incoming particles to their orientations, we may define these functions as follows:
\begin{align}
S^3\ni{\mathscr A}({\mathbf a},{\mathbf s}^i_1)\,&=\lim_{{\mathbf s}_1\,\rightarrow\,\mu_1{\mathbf a}}\left\{-\,{\mathbf D}({\mathbf a})\,{\mathbf L}({\mathbf s}_1)\right\} \label{79-nmn} \\
&=\lim_{{\mathbf s}_1\,\rightarrow\,\mu_1{\mathbf a}}\left\{-(I_3{\mathbf a})(I_3{\mathbf s}_1)\right\} \\
&=\lim_{{\mathbf s}_1\,\rightarrow\,\mu_1{\mathbf a}}\left\{{\mathbf a}\cdot{\mathbf s}_1+I_3({\mathbf a}\times{\mathbf s}_1)\right\} \\
&=\lim_{{\mathbf s}_1\,\rightarrow\,\mu_1{\mathbf a}}\left\{\cos(\eta_{{\mathbf a}{\mathbf s}_1})+(I_3{\mathbf r}_{1})\sin(\eta_{{\mathbf a}{\mathbf s}_1})\right\} \label{20no}\\
&=\lim_{{\mathbf s}_1\,\rightarrow\,\mu_1{\mathbf a}}\left\{\,+\,{\mathbf q}(\eta_{{\mathbf a}{\mathbf s}_1},\,{\mathbf r}_1)\right\}, \label{84-nom}
\end{align}
where $\eta_{{\mathbf a}{\mathbf s}_1}$ is the angle between the spin direction ${\mathbf s}_1$ and the detector direction ${\mathbf a}$ chosen by Alice as depicted in Fig.~\ref{Fig-1}, and ${\mathbf r}_1=\frac{{\mathbf a}\times{\mathbf s}_1}{||{\mathbf a}\times{\mathbf s}_1||}$ is the normalized rotation axis of the quaternion ${\mathbf q}(\eta_{{\mathbf a}{\mathbf s}_1},\,{\mathbf r}_1)$ in $S^3$. Thus, given the definition (\ref{79-nom}) of $\mu_1$ as the function $\text{sign}(\mathbf{a}\cdot\mathbf{s}^i_1)$, the spin direction ${\mathbf s}_1$, which mathematically acts as a dummy variable in the above limits while approaching the detector ${\mathbf D}({\mathbf a})$, will tend to $+{\mathbf a}$ if initially the two unit vectors ${\mathbf a}$ and ${\mathbf{s}^i_1}$ happen to be pointing through the same hemisphere of $S^2$ centered at the origin of ${\mathbf{s}^i_1}$, and otherwise the spin direction ${\mathbf s}_1$ will tend to $-{\mathbf a}$. Consequently, as ${\mathbf s}_1\rightarrow\mu_1{\mathbf a}=\pm{\mathbf a}$ during the detection process by Alice so that the angle $\eta_{{\mathbf a}{\mathbf s}_1}\!\rightarrow0$ or $\pi$, the binary value of the result observed by her is obtained because $\cos(\eta_{{\mathbf a}{\mathbf s}_1})\rightarrow \pm1$ and $\sin(\eta_{{\mathbf a}{\mathbf s}_1})\rightarrow 0$, giving
\begin{equation}
{\mathscr A}({\mathbf a},{\mathbf s}^i_1)\longrightarrow+\mu_1=\pm1. \label{85-nom}
\end{equation}
Thus, using a limit function to model the detection process of the spin values allows us to retain their directional or contextual information passively, without having to explicitly use directed numbers such as vectors or bivectors to represent scalar-valued measurement results, $\pm1$, observed in the Bell-test experiments.

Similarly, the measurement function for Bob is defined as
\begin{align}
S^3\ni{\mathscr B}({\mathbf b},{\mathbf s}^i_2)\,&=\lim_{{\mathbf s}_2\,\rightarrow\,\mu_2{\mathbf b}}\left\{+\,{\mathbf L}({\mathbf s}_2)\,{\mathbf D}({\mathbf b})\right\} \label{85-nmn} \\
&=\lim_{{\mathbf s}_2\,\rightarrow\,\mu_2{\mathbf b}}\left\{+(I_3{\mathbf s}_2)(I_3{\mathbf b})\right\} \\
&=\lim_{{\mathbf s}_2\,\rightarrow\,\mu_2{\mathbf b}}\left\{-\,{\mathbf s}_2\cdot{\mathbf b}-I_3({\mathbf s}_2\times{\mathbf b})\right\} \\
&=\lim_{{\mathbf s}_2\,\rightarrow\,\mu_2{\mathbf b}}\left\{-\cos(\eta_{{\mathbf s}_2{\mathbf b}})-(I_3{\mathbf r}_{2})\sin(\eta_{{\mathbf s}_2{\mathbf b}})\right\} \\
&=\lim_{{\mathbf s}_2\,\rightarrow\,\mu_2{\mathbf b}}\left\{\,-\,{\mathbf q}(\eta_{{\mathbf s}_2{\mathbf b}},\,{\mathbf r}_2)\right\}, \label{90-nom}
\end{align}
where $\eta_{{\mathbf s}_2{\mathbf b}}$ is the angle between the spin direction ${\mathbf s}_2$ and the detector direction ${\mathbf b}$ chosen by Bob as depicted in Fig.~\ref{Fig-1}, and ${\mathbf r}_2=\frac{{\mathbf s}_2\times{\mathbf b}}{||{\mathbf s}_2\times{\mathbf b}||}$ is the normalized rotation axis of the quaternion ${\mathbf q}(\eta_{{\mathbf s}_2{\mathbf b}},\,{\mathbf r}_2)$ in $S^3$. Consequently, analogous to Alice's detection process, as ${\mathbf s}_2\rightarrow\mu_2{\mathbf b}=\pm{\mathbf b}$ during the detection process by Bob so that the angle $\eta_{{\mathbf s}_2{\mathbf b}}\!\rightarrow0$ or $\pi$, the binary value of the result observed by him is obtained because $\cos(\eta_{{\mathbf s}_2{\mathbf b}})\rightarrow \pm1$ and $\sin(\eta_{{\mathbf s}_2{\mathbf b}})\rightarrow 0$, giving
\begin{equation}
{\mathscr B}({\mathbf b},{\mathbf s}^i_2)\longrightarrow-\mu_2=\mp1. \label{91-nom}
\end{equation}

Recall that the spin variables in the EPR-Bohm experiments do not change until detection. Their initial spin directions and rotation senses acquired from the source remain the same until the spins reach the detectors, because of the conservation of the initial zero spin angular momentum. The limiting process appearing in the definition (\ref{79-nmn}) of the measurement function ${{\mathscr A}({\bf a},\,{\bf s}^i_1)}$ thus physically describes the process of detection of the spin ${\bf L}({\bf s}_1)$ by the detector ${\bf D}({\bf a})$, analogous to what occurs in a Stern–Gerlach device, whereas mathematically the limiting process is a Hopf map $h:S^3\rightarrow{S^2}$ that projects the Hopf circle $S^1$ seen in (\ref{20no}) onto a point $\mu_1=\pm1$ on the $S^2\hookrightarrow\mathrm{I\!R}^3$ assumed by Bell \cite{Bell-1964} (for further details see Appendix~\ref{QandA}). 

Now, it follows from (\ref{85-nom}) and (\ref{91-nom}) that, in general, for the choices ${\mathbf a}\not={\mathbf b}$, the product of the results observed by Alice and Bob will fluctuate between $-1$ and $+1$:
\begin{align}
{\mathscr A}&({\mathbf a},{\mathbf s}^i_1)\,{\mathscr B}({\mathbf b},{\mathbf s}^i_2) \notag \\
&=\left[\lim_{{\mathbf s}_1\,\rightarrow\,\mu_1{\mathbf a}}\left\{-\,{\mathbf D}({\mathbf a})\,{\mathbf L}({\mathbf s}_1)\right\}\right]\left[\lim_{{\mathbf s}_2\,\rightarrow\,\mu_2{\mathbf b}}\left\{+\,{\mathbf L}({\mathbf s}_2)\,{\mathbf D}({\mathbf b})\right\}\right] \notag \\
&= -\mu_1\mu_2=-\,(\pm1)(\pm1)=\mp1. \label{92-nom}
\end{align}
On the other hand, for the choice ${\mathbf b}={\mathbf a}$, we have ${\mu}_2={\mu}_1$ because ${\mathbf s}^i_2=\,{\mathbf s}^i_1$, and perfect anti-correlations are predicted:
\begin{align}
{\mathscr A}&({\mathbf a},{\mathbf s}^i_1)\,{\mathscr B}({\mathbf a},{\mathbf s}^i_2) \notag \\
&=\left[\lim_{{\mathbf s}_1\,\rightarrow\,\mu_1{\mathbf a}}\left\{-\,{\mathbf D}({\mathbf a})\,{\mathbf L}({\mathbf s}_1)\right\}\right]\left[\lim_{{\mathbf s}_2\,\rightarrow\,\mu_1{\mathbf a}}\left\{+\,{\mathbf L}({\mathbf s}_2)\,{\mathbf D}({\mathbf a})\right\}\right] \notag \\
&=-(\mu_1)^2=-1.
\end{align}

It is important to note that the functions ${\mathscr A}({\mathbf a},{\mathbf s}^i_1)$ and ${\mathscr B}({\mathbf b},{\mathbf s}^i_2)$ defined in (\ref{79-nmn}) and (\ref{85-nmn}) are  manifestly local-realistic in the sense espoused by Einstein and formalized by Bell in \cite{Bell-1964}. Apart from the hidden variable ${\bf s}^i_1$, the result ${{\mathscr A}=\pm1}$ depends {\it only} on the measurement direction ${\bf a}$, chosen freely by Alice, regardless of Bob's actions. And, analogously, apart from the hidden variable ${\bf s}^i_2$, the result ${{\mathscr B}=\pm1}$ depends {\it only} on the measurement direction ${\bf b}$, chosen freely by Bob, regardless of Alice's actions. In particular, the function ${{\mathscr A}({\bf a},\,{\bf s}^i_1)}$ {\it does not} depend on ${\bf b}$ or ${\mathscr B}$ and the function ${{\mathscr B}({\bf b},\,{\bf s}^i_2)}$ {\it does not} depend on ${\bf a}$ or ${\mathscr A}$. Moreover, the hidden variables ${\bf s}^i_1$ and ${\bf s}^i_2$ do not depend on ${\bf a}$, ${\bf b}$, ${\mathscr A}$, or ${\mathscr B}$ (see also Appendix~\ref{QandA}).

\section{Computing the singlet correlations within 3-sphere} \label{Sec-3}

It is also important to note that the conservation of the initial zero spin angular momentum (\ref{55}) requires the equality ${\mathbf s}_1=\,{\mathbf s}_2$ to hold throughout the free evolution of the spins, but not necessarily during their detection processes \cite{IEEE-2}. Consequently, while the physical interactions are taking place during the independent detection processes at the spacelike separated observation stations, ${\mathbf s}_2$ can tend to $\mu_2{\mathbf b}$ jointly as ${\mathbf s}_1$ tends to $\mu_1{\mathbf a}$, despite the fact that ${\mathbf s}_1=\,{\mathbf s}_2$ must hold during the free evolution of the two spins. If now the initial direction ${\mathbf s}^i={\mathbf s}^i_1={\mathbf s}^i_2$ of the spins originating at the source is assumed to be uniformly distributed over $S^2$ as in Bell's local model \cite{Bell-1964,Peres}, specified by a normalized probability measure $p({\mathbf s}^i)$ so that
\begin{equation}
\int_{S^2}p({\mathbf s}^i)\;d{\mathbf s}^i=1,
\end{equation}
then the results (\ref{85-nom}) and (\ref{91-nom}) independently observed by Alice and Bob will be equal to $+1$ or $-1$ with 50/50 chance. Consequently, in analogy with the quantum mechanical predictions, the expectation values of these results will vanish:
\begin{equation}
{\cal E}(\mathbf{a})=\int_{S^2} \mathscr{A}({\mathbf a},\mathbf{s}^i_1)\,p({\mathbf s}^i)\,d{\mathbf s}^i = 0
\end{equation}
and
\begin{equation}
{\cal E}(\mathbf{b})=\int_{S^2}\mathscr{B}({\mathbf b},\mathbf{s}^i_2)\,p({\mathbf s}^i)\,d{\mathbf s}^i=0,
\end{equation}
where the subscripts 1 and 2 on the spin directions are retained for clarity even though they are the same direction.

The question now is: What will be the correlations within $S^3$ between the results ${\mathscr A}({\mathbf a},{\mathbf s}^i_1)$ and ${\mathscr B}({\mathbf b},{\mathbf s}^i_2)$ observed jointly but independently by Alice and Bob, in coincident counts, at a spacelike distance from each other? To answer this question, recall that the standard function for computing correlations as expectation values employed by Bell in \cite{Bell-1964} is
\begin{equation}
{\cal E}({\mathbf a},{\mathbf b})=\int_{S^2}{\mathscr A}({\mathbf a},{\mathbf s}^i_1)\;{\mathscr B}({\mathbf b},{\mathbf s}^i_2)\;p({\mathbf s}^i)\,d{\mathbf s}^i. \label{for-1}
\end{equation}
This expectation function respects two important properties. First, it is a scalar-valued function: ${\cal E}({\mathscr A},{\mathscr B})\!\!: \mathrm{I\!R}\times\mathrm{I\!R}\longrightarrow\mathrm{I\!R}$. Second, it is invariant under the change of order of the product of the results ${\mathscr A}({\mathbf a},{\mathbf s}^i_1)$ and ${\mathscr B}({\mathbf b},{\mathbf s}^i_2)$. In other words, it is a symmetric function:  ${\cal E}({\mathscr A},{\mathscr B})={\cal E}({\mathscr B},{\mathscr A})$. These properties are trivially satisfied in Bell's local model set within $\mathrm{I\!R}^3$, because in it the results ${\mathscr A}({\mathbf a},{\mathbf s}^i_1)$ and ${\mathscr B}({\mathbf b},{\mathbf s}^i_2)$ are scalar-valued functions by construction, and therefore commute with each other. With some care, these two properties of the expectation function can be satisfied also within our $S^3$ model despite the fact that the measurement functions in our model are composed of products of bivectors, or quaternions, that do not commute in general:
\begin{align}
{\mathscr A}&({\mathbf a},{\mathbf s}^i_1)\,{\mathscr B}({\mathbf b},{\mathbf s}^i_2) \notag \\
&=\left[\lim_{{\mathbf s}_1\,\rightarrow\,\mu_1{\mathbf a}}\left\{-\,{\mathbf D}({\mathbf a})\,{\mathbf L}({\mathbf s}_1)\right\}\right]\left[\lim_{{\mathbf s}_2\,\rightarrow\,\mu_2{\mathbf b}}\left\{+\,{\mathbf L}({\mathbf s}_2)\,{\mathbf D}({\mathbf b})\right\}\right] \notag \\
&=\lim_{\substack{{\mathbf s}_1\,\rightarrow\,\mu_1{\mathbf a} \\ {\mathbf s}_2\,\rightarrow\,\mu_2{\mathbf b}}}\Big\{-{\mathbf D}({\mathbf a})\,{\mathbf L}({\mathbf s}_1)\,{\mathbf L}({\mathbf s}_2)\,{\mathbf D}({\mathbf b})\Big\}, \label{36-unm}
\end{align}
but
\begin{align}
{\mathscr B}&({\mathbf b},{\mathbf s}^i_2)\,{\mathscr A}({\mathbf a},{\mathbf s}^i_1) \notag \\
&=\left[\lim_{{\mathbf s}_2\,\rightarrow\,\mu_2{\mathbf b}}\left\{+\,{\mathbf L}({\mathbf s}_2)\,{\mathbf D}({\mathbf b})\right\}\right]\left[\lim_{{\mathbf s}_1\,\rightarrow\,\mu_1{\mathbf a}}\left\{-\,{\mathbf D}({\mathbf a})\,{\mathbf L}({\mathbf s}_1)\right\}\right] \notag \\
&=\lim_{\substack{{\mathbf s}_2\,\rightarrow\,\mu_2{\mathbf b} \\ {\mathbf s}_1\,\rightarrow\,\mu_1{\mathbf a}}}\Big\{-{\mathbf L}({\mathbf s}_2)\,{\mathbf D}({\mathbf b})\,{\mathbf D}({\mathbf a})\,{\mathbf L}({\mathbf s}_1)\Big\}. \label{37-rel}
\end{align}
Here we have used the ``product of limits equal to limits of product'' rule. Now, a very important property of the 3-sphere defined in (\ref{nonsin}) is that it remains closed under multiplications of its constituent quaternions. Therefore, the product of the unit quaternions ${\mathbf q}(\eta_{{\mathbf a}{\mathbf s}_1},\,{\mathbf r}_1)\in S^3$ and ${\mathbf q}(\eta_{{\mathbf s}_2{\mathbf b}},\,{\mathbf r}_2)\in S^3$ appearing in the definitions (\ref{84-nom}) and (\ref{90-nom}) of the measurement outcomes,
\begin{equation}
{\mathbf q}(\eta_{{\mathbf a}{\mathbf s}_1},\,{\mathbf r}_1)\,{\mathbf q}(\eta_{{\mathbf s}_2{\mathbf b}},\,{\mathbf r}_2)={\mathbf q}(\eta_{{\mathbf u}{\mathbf v}},\,{\mathbf r}_0),
\end{equation}
is also a unit quaternion in $S^3$ (while we will not need explicit expression of ${\mathbf q}(\eta_{{\mathbf u}{\mathbf v}},\,{\mathbf r}_0)$ here, its derivation can be found in appendix~A.5 of \cite{local}). As a result, the following equalities hold:
\begin{align}
\Big[\lim_{{\mathbf s}_1\,\rightarrow\,\mu_1{\mathbf a}}\big\{+{\mathbf q}&(\eta_{{\mathbf a}{\mathbf s}_1},\,{\mathbf r}_1)\big\}\Big]\Big[\lim_{{\mathbf s}_2\,\rightarrow\,\mu_2{\mathbf b}}\big\{-\,{\mathbf q}(\eta_{{\mathbf s}_2{\mathbf b}},\,{\mathbf r}_2)\big\}\Big] \notag \\ 
&=\lim_{\substack{{\mathbf s}_1\,\rightarrow\,\mu_1{\mathbf a} \\ {\mathbf s}_2\,\rightarrow\,\mu_2{\mathbf b}}}\big\{-{\mathbf q}(\eta_{{\mathbf a}{\mathbf s}_1},\,{\mathbf r}_1)\,{\mathbf q}(\eta_{{\mathbf s}_2{\mathbf b}},\,{\mathbf r}_2)\big\} \\
&=\lim_{\substack{{\mathbf s}_1\,\rightarrow\,\mu_1{\mathbf a} \\ {\mathbf s}_2\,\rightarrow\,\mu_2{\mathbf b}}}\big\{-{\mathbf q}(\eta_{{\mathbf u}{\mathbf v}},\,{\mathbf r}_0)\big\} \\
&\equiv\lim_{\substack{{\mathbf s}_1\,\rightarrow\,\mu_1{\mathbf a} \\ {\mathbf s}_2\,\rightarrow\,\mu_2{\mathbf b}}}\Big\{-{\mathbf D}({\mathbf a})\,{\mathbf L}({\mathbf s}_1)\,{\mathbf L}({\mathbf s}_2)\,{\mathbf D}({\mathbf b})\Big\}. \label{limitproduct}
\end{align}
Thus, the products such as ${\mathscr A}({\mathbf a},{\mathbf s}^i_1)\,{\mathscr B}({\mathbf b},{\mathbf s}^i_2)$ of measurement outcomes are also limiting scalar points of some quaternions within $S^3$. However, it is easy to recognize from equations (\ref{36-unm}) and (\ref{37-rel}) that, since bivectors do not commute in general, the products ${\mathscr A}({\mathbf a},{\mathbf s}^i_1)\,{\mathscr B}({\mathbf b},{\mathbf s}^i_2)$ and ${\mathscr B}({\mathbf b},{\mathbf s}^i_2)\,{\mathscr A}({\mathbf a},{\mathbf s}^i_1)$ are limiting scalar points of two entirely {\it different} quaternions within $S^3$. Moreover, the symmetry of the experimental setup shown in Fig.~\ref{Fig-1} necessitates that the product of the results ${\mathscr A}({\mathbf a},{\mathbf s}^i_1)$ and ${\mathscr B}({\mathbf b},{\mathbf s}^i_2)$ would be equally likely to be a limiting scalar value of the quaternion $-{\mathbf D}({\mathbf a})\,{\mathbf L}({\mathbf s}_1)\,{\mathbf L}({\mathbf s}_2)\,{\mathbf D}({\mathbf b})$ as it would be a limiting scalar value of the quaternion $-{\mathbf L}({\mathbf s}_2)\,{\mathbf D}({\mathbf b})\,{\mathbf D}({\mathbf a})\,{\mathbf L}({\mathbf s}_1)$. Therefore, since the expectation value is essentially the average value of the product of the results ${\mathscr A}({\mathbf a},{\mathbf s}^i_1)$ and ${\mathscr B}({\mathbf b},{\mathbf s}^i_2)$, we must average over {\it both} orders of the product with equal weight to respect the defining properties of the expectation function we discussed above, by writing (\ref{for-1}) in the symmetric form\footnote{Incidentally, in the model presented in \cite{RSOS,IEEE-1,IEEE-2} the symmetric form (\ref{for-2}) of the expectation function is automatically achieved by the choice of an orientation $\lambda=\pm1$ of $S^3$ as a hidden variable (cf. footnote~\ref{foot-2}). Recall that, unlike $S^2$ whose tangent bundle is non-trivial, $TS^2\not= S^2\times{\mathrm{I\!R}}^2$, the quaternionic 3-sphere, or $S^3$, is an {\it orientable} manifold whose tangent bundle is trivial, $TS^3= S^3\times{\mathrm{I\!R}}^3$, as we noted in (\ref{cfeq3}). That is to say, unlike on $S^2$, a consistent sense of orientation or handedness can be assigned to each point on $S^3$, so that ``hair'' on $S^3$ can be ``combed'' without creating a ``cowlick.'' Thus, one can consistently assign either a right-handed sense of rotation or a left-handed sense of rotation at each point on $S^3$. This allows us to identify the orientation $\lambda=\pm1$ of $S^3$ as a hidden variable with a 50/50 chance of either orientation being initially realized, which restores the symmetry in the expectation function (\ref{for-1}), because a flip in the orientation induces a flip in the order of the product of any two quaternions, such as those seen in the products (\ref{36-unm}) and (\ref{37-rel}).\label{foot-3}} as 
\begin{align}
{\cal E}({\mathbf a},{\mathbf b})=\int_{S^2}\Big[\frac{1}{2}&\Big\{{\mathscr A}({\mathbf a},{\mathbf s}^i_1)\,{\mathscr B}({\mathbf b},{\mathbf s}^i_2) \notag \\
&+{\mathscr B}({\mathbf b},{\mathbf s}^i_2)\,{\mathscr A}({\mathbf a},{\mathbf s}^i_1)\Big\}\Big]p({\mathbf s}^i)\,d{\mathbf s}^i. \label{for-2}
\end{align}
The statistical properties of the expectation functions (\ref{for-1}) and (\ref{for-2}) are identical, however, because of the arithmetical identity
\begin{align}
{\mathscr A}({\mathbf a},{\mathbf s}^i_1)&\,{\mathscr B}({\mathbf b},{\mathbf s}^i_2) \notag \\ 
&\equiv \frac{1}{2}\Big\{{\mathscr A}({\mathbf a},{\mathbf s}^i_1)\,{\mathscr B}({\mathbf b},{\mathbf s}^i_2) + {\mathscr B}({\mathbf b},{\mathbf s}^i_2)\,{\mathscr A}({\mathbf a},{\mathbf s}^i_1)\Big\}.
\end{align}
Indeed, in analogy with (\ref{92-nom}) it is easy to verify that
\begin{align}
{\mathscr B}&({\mathbf b},{\mathbf s}^i_2)\,{\mathscr A}({\mathbf a},{\mathbf s}^i_1) \notag \\
&=\left[\lim_{{\mathbf s}_2\,\rightarrow\,\mu_2{\mathbf b}}\left\{+\,{\mathbf L}({\mathbf s}_2)\,{\mathbf D}({\mathbf b})\right\}\right]\!\left[\lim_{{\mathbf s}_1\,\rightarrow\,\mu_1{\mathbf a}}\left\{-\,{\mathbf D}({\mathbf a})\,{\mathbf L}({\mathbf s}_1)\right\}\right] \notag \\
&= -\mu_2\mu_1=-\mu_1\mu_2=\mp1,
\end{align}
which follows from the definitions (\ref{79-nmn}) and (\ref{85-nmn}). Consequently, the equivalence of formulae (\ref{for-1}) and (\ref{for-2}) follows: 
\begin{align}
\frac{1}{2}\Big\{{\mathscr A}({\mathbf a},{\mathbf s}^i_1)\;{\mathscr B}({\mathbf b},{\mathbf s}^i_2) + {\mathscr B}&({\mathbf b},{\mathbf s}^i_2)\;{\mathscr A}({\mathbf a},{\mathbf s}^i_1)\Big\} \notag \\
&=\frac{1}{2}\Big\{-\mu_1\mu_2 - \mu_2\mu_1\,\Big\} \notag \\
&=-\mu_1\mu_2 = \mp1. \label{equabell}
\end{align}

Now, the equality (\ref{37-rel}) can be simplified by multiplying its right-hand side with ${\mathbf L}({\mathbf s}_2)$ from the left and from the right in its numerator and denominator to give
\begin{align}
{\mathscr B}&({\mathbf b},{\mathbf s}^i_2)\,{\mathscr A}({\mathbf a},{\mathbf s}^i_1) \notag \\
&=\lim_{\substack{{\mathbf s}_2\,\rightarrow\,\mu_2{\mathbf b} \\ {\mathbf s}_1\,\rightarrow\,\mu_1{\mathbf a}}}\left\{-\frac{{\mathbf L}({\mathbf s}_2)\,{\mathbf L}({\mathbf s}_2)\,{\mathbf D}({\mathbf b})\,{\mathbf D}({\mathbf a})\,{\mathbf L}({\mathbf s}_1)\,{\mathbf L}({\mathbf s}_2)}{{\mathbf L}({\mathbf s}_2)\,{\mathbf L}({\mathbf s}_2)}\right\} \\
&=\lim_{\substack{{\mathbf s}_2\,\rightarrow\,\mu_2{\mathbf b} \\ {\mathbf s}_1\,\rightarrow\,\mu_1{\mathbf a}}}\Big\{-{\mathbf D}({\mathbf b})\,{\mathbf D}({\mathbf a})\,{\mathbf L}({\mathbf s}_1)\,{\mathbf L}({\mathbf s}_2)\Big\}, \label{39-opp}
\end{align}
where in the last step we have used the fact that bivectors square to $-1$: ${\bf L}({\bf s}_2)\,{\bf L}({\bf s}_2)=\,I_3{\bf s}_2\,I_3{\bf s}_2=(I_3)^2\,{\bf s}_2\,{\bf s}_2=-1.$ Consequently, using the equality (\ref{39-opp}), we see that the limit relation (\ref{37-rel}) reduces to the following convenient form:
\begin{align}
&{\mathscr B}({\mathbf b},{\mathbf s}^i_2)\,{\mathscr A}({\mathbf a},{\mathbf s}^i_1) \notag \\
&=\left[\lim_{{\mathbf s}_2\,\rightarrow\,\mu_2{\mathbf b}}\left\{+\,{\mathbf L}({\mathbf s}_2)\,{\mathbf D}({\mathbf b})\right\}\right]\left[\lim_{{\mathbf s}_1\,\rightarrow\,\mu_1{\mathbf a}}\left\{-\,{\mathbf D}({\mathbf a})\,{\mathbf L}({\mathbf s}_1)\right\}\right] \notag \\
&=\lim_{\substack{{\mathbf s}_2\,\rightarrow\,\mu_2{\mathbf b} \\ {\mathbf s}_1\,\rightarrow\,\mu_1{\mathbf a}}}\Big\{-{\mathbf D}({\mathbf b})\,{\mathbf D}({\mathbf a})\,{\mathbf L}({\mathbf s}_1)\,{\mathbf L}({\mathbf s}_2)\Big\}. \label{40-rtc}
\end{align}
The equalities (\ref{36-unm}) and (\ref{40-rtc}) are easy to verify by taking limits explicitly on both sides. Nevertheless, for completeness of our derivation, we have proved them in Appendix~\ref{appa} below. With them, we can now derive the correlations using
\begin{align}
&{\cal E}({\mathbf a},{\mathbf b}) \notag \\
&=\!\!\!\!\int_{S^2}\frac{1}{2}\Big[{\mathscr A}({\mathbf a},{\mathbf s}^i_1)\;{\mathscr B}({\mathbf b},{\mathbf s}^i_2)+{\mathscr B}({\mathbf b},{\mathbf s}^i_2)\;{\mathscr A}({\mathbf a},{\mathbf s}^i_1)\Big]\,p({\mathbf s}^i)\,d{\mathbf s}^i \notag \\
&=\!\!\!\!\int_{S^2}\frac{1}{2}\Bigg[\lim_{\substack{{\mathbf s}_1\,\rightarrow\,\mu_1{\mathbf a} \\ {\mathbf s}_2\,\rightarrow\,\mu_2{\mathbf b}}}\Big\{-{\mathbf D}({\mathbf a})\,{\mathbf L}({\mathbf s}_1)\,{\mathbf L}({\mathbf s}_2)\,{\mathbf D}({\mathbf b})\Big\}\, \notag \\
&\;\;\;\;\;\;\;\;\;\;\;\;+\!\lim_{\substack{{\mathbf s}_2\,\rightarrow\,\mu_2{\mathbf b} \\ {\mathbf s}_1\,\rightarrow\,\mu_1{\mathbf a}}}\Big\{-{\mathbf D}({\mathbf b})\,{\mathbf D}({\mathbf a})\,{\mathbf L}({\mathbf s}_1)\,{\mathbf L}({\mathbf s}_2)\Big\}\Bigg]p({\mathbf s}^i)\,d{\mathbf s}^i.
\end{align}
Now, before proceeding further we have to make a choice. Either we ignore the fact that the spin angular momentum is conserved during the free evaluation of the spins (cf. Fig.~\ref{Fig-1}), which amounts to ignoring the conditions (\ref{55}) and (\ref{56}), or set ${\mathbf L}({\mathbf s}_1)\,{\mathbf L}({\mathbf s}_2)=-1$ respecting the conservation law. We first proceed by ignoring this condition, which leads to
\begin{widetext}
\begin{align}
{\cal E}({\mathbf a},{\mathbf b})&=\int_{S^2}\frac{1}{2}\Bigg[\lim_{\substack{{\mathbf s}_1\,\rightarrow\,\mu_1{\mathbf a} \\ {\mathbf s}_2\,\rightarrow\,\mu_2{\mathbf b}}}\Big\{-{\mathbf D}({\mathbf a})\,{\mathbf L}({\mathbf s}_1)\,{\mathbf L}({\mathbf s}_2)\,{\mathbf D}({\mathbf b})\Big\}\,+\lim_{\substack{{\mathbf s}_2\,\rightarrow\,\mu_2{\mathbf b} \\ {\mathbf s}_1\,\rightarrow\,\mu_1{\mathbf a}}}\Big\{-{\mathbf D}({\mathbf b})\,{\mathbf D}({\mathbf a})\,{\mathbf L}({\mathbf s}_1)\,{\mathbf L}({\mathbf s}_2)\Big\}\Bigg]p({\mathbf s}^i)\,d{\mathbf s}^i \label{43-say} \\
&=\int_{S^2}\frac{1}{2}\left[\Big\{-\mu_1\mu_2\,{\mathbf D}({\mathbf a})\,{\mathbf L}({\mathbf a})\,{\mathbf L}({\mathbf b})\,{\mathbf D}({\mathbf b})\Big\}\,+\Big\{-\mu_1\mu_2\,{\mathbf D}({\mathbf b})\,{\mathbf D}({\mathbf a})\,{\mathbf L}({\mathbf a})\,{\mathbf L}({\mathbf b})\Big\}\right]p({\mathbf s}^i)\,d{\mathbf s}^i \\
&=\int_{S^2}\frac{1}{2}\left[\Big\{-\mu_1\mu_2\,I_3{\mathbf a}\,I_3{\mathbf a}\,I_3{\mathbf b}\,I_3{\mathbf b}\Big\}\,+\Big\{-\mu_1\mu_2\,I_3{\mathbf b}\,I_3{\mathbf a}\,I_3{\mathbf a}\,I_3{\mathbf b}\Big\}\right]p({\mathbf s}^i)\,d{\mathbf s}^i \\
&=\int_{S^2}\frac{1}{2}\left[\Big\{-\mu_1\mu_2\,(I_3{\mathbf a})^2\,(I_3{\mathbf b})^2\Big\}\,+\Big\{-\mu_1\mu_2\,(I_3{\mathbf b})\,(I_3{\mathbf a})^2\,(I_3{\mathbf b})\Big\}\right]p({\mathbf s}^i)\,d{\mathbf s}^i \label{46-hose} \\
&=\int_{S^2}\frac{1}{2}\left[\Big\{-\mu_1\mu_2\,\Big\}\,+\Big\{-\mu_1\mu_2\,\Big\}\right]p({\mathbf s}^i)\,d{\mathbf s}^i \label{47-hose}\\
&=\int_{S^2}\Big\{-\mu_1\mu_2\,\Big\}\;p({\mathbf s}^i)\,d{\mathbf s}^i \\
&=\int_{S^2}\Big\{\!-\text{sign}(\mathbf{a}\cdot\mathbf{s}^i_1)\,\text{sign}(\mathbf{b}\cdot\mathbf{s}^i_2)\Big\}\;p({\mathbf s}^i)\;d{\mathbf s}^i, \label{50-int}
\end{align}
\end{widetext}
where in the step from (\ref{46-hose}) to (\ref{47-hose}) we have used $(I_3{\mathbf a})^2=(I_3{\mathbf b})^2=-1$, and in the last step the expressions for $\mu_1$ and $\mu_2$ are substituted from (\ref{79-nom}). Needless to say, we could have inferred the last two steps immediately from (\ref{92-nom}), but have preferred to carry out longer calculation to demonstrate that our derivation is consistent and reproduces the known result. Indeed, the integral in (\ref{50-int}) is identical to the one that appears in Bell's local model \cite{Bell-1964,Peres}. However, both Bell and Peres claim that it leads to the following weak correlations that are incapable of exceeding the bounds set by Bell inequalities:
\begin{align}
{\cal E}({\mathbf a},{\mathbf b})\,&=\int_{S^2}\!\left\{-\,\text{sign}(\mathbf{a}\cdot\mathbf{s}^i_1)\;\text{sign}(\mathbf{b}\cdot\mathbf{s}^i_2)\right\}\,p({\mathbf s}^i)\;d{\mathbf s}^i \notag \\
&=\,
\begin{cases}
-\,1\,+\,\frac{2}{\pi}\,\eta_{{\mathbf a}{\mathbf b}}
\;\;\;\text{if} &\!\! 0 \leqslant \eta_{{\mathbf a}{\mathbf b}} \leqslant \pi, \\
+\,3\,-\,\frac{2}{\pi}\,\eta_{{\mathbf a}{\mathbf b}}
\;\;\;\text{if} &\!\! \pi \leqslant \eta_{{\mathbf a}{\mathbf b}} \leqslant 2\pi.
\end{cases} \label{51-con}
\end{align}
In Fig.~\ref{Fig-2} these weak correlations are depicted by saw-tooth shaped dashed straight lines. By contrast, the solid cosine curve in Fig.~\ref{Fig-2} depicts the strong singlet correlations predicted by our 3-sphere model as well as quantum mechanics. They follow if the spin angular momentum is assumed to be conserved during the free evolution of the constituent spins by setting ${\mathbf L}({\mathbf s}_1)\,{\mathbf L}({\mathbf s}_2)=-1$ in the above derivation of the correlations, as we now proceed to demonstrate:

\begin{widetext}
\begin{align}
{\cal E}({\mathbf a},{\mathbf b})&=\int_{S^2}\frac{1}{2}\Big[{\mathscr A}({\mathbf a},{\mathbf s}^i_1)\;{\mathscr B}({\mathbf b},{\mathbf s}^i_2)\,+\,{\mathscr B}({\mathbf b},{\mathbf s}^i_2)\;{\mathscr A}({\mathbf a},{\mathbf s}^i_1)\Big]\,p({\mathbf s}^i)\,d{\mathbf s}^i \label{52-step} \\
&=\int_{S^2}\frac{1}{2}\Bigg[\lim_{\substack{{\mathbf s}_1\,\rightarrow\,\mu_1{\mathbf a} \\ {\mathbf s}_2\,\rightarrow\,\mu_2{\mathbf b}}}\Big\{-{\mathbf D}({\mathbf a})\,{\mathbf L}({\mathbf s}_1)\,{\mathbf L}({\mathbf s}_2)\,{\mathbf D}({\mathbf b})\Big\}+\!\lim_{\substack{{\mathbf s}_2\,\rightarrow\,\mu_2{\mathbf b} \\ {\mathbf s}_1\,\rightarrow\,\mu_1{\mathbf a}}}\Big\{-{\mathbf D}({\mathbf b})\,{\mathbf D}({\mathbf a})\,{\mathbf L}({\mathbf s}_1)\,{\mathbf L}({\mathbf s}_2)\Big\}\Bigg]p({\mathbf s}^i)\,d{\mathbf s}^i \label{53-step} \\
&=\int_{S^2}\frac{1}{2}\left[\lim_{\substack{{\mathbf s}_1\,\rightarrow\,\mu_1{\mathbf a} \\ {\mathbf s}_2\,\rightarrow\,\mu_2{\mathbf b}}}\Big\{\,{\mathbf D}({\mathbf a})\,{\mathbf D}({\mathbf b})\Big\}+\!\lim_{\substack{{\mathbf s}_2\,\rightarrow\,\mu_2{\mathbf b} \\ {\mathbf s}_1\,\rightarrow\,\mu_1{\mathbf a}}}\Big\{\,{\mathbf D}({\mathbf b})\,{\mathbf D}({\mathbf a})\Big\}\right]p({\mathbf s}^i)\,d{\mathbf s}^i \label{54-step} \\
&=\int_{S^2}\frac{1}{2}\left[\Big\{\,I_3{\mathbf a}\,I_3{\mathbf b}\Big\}\,+\Big\{\,I_3{\mathbf b}\,I_3{\mathbf a}\Big\}\right]p({\mathbf s}^i)\,d{\mathbf s}^i \\ 
&=\frac{1}{2}\left[\Big\{\,(I_3)^2\,{\mathbf a}\,{\mathbf b}\Big\}\,+\Big\{\,(I_3)^2\,{\mathbf b}\,{\mathbf a}\Big\}\right]\int_{S^2}p({\mathbf s}^i)\,d{\mathbf s}^i \label{sym}\\
&=-\frac{1}{2}\big[\,{\mathbf a}\,{\mathbf b}+{\mathbf b}\,{\mathbf a}\,\big] \label{pro}\\
&=-{\mathbf a}\cdot{\mathbf b}\equiv\,-\cos(\,\eta_{{\mathbf a}{\mathbf b}}), \label{whichr}
\end{align}
\end{widetext}
where the step (\ref{54-step}) follows from the step (\ref{53-step}) by setting ${\mathbf L}({\mathbf s}_1)\,{\mathbf L}({\mathbf s}_2)=-1$ in (\ref{53-step}), as required by the conservation of initial zero spin angular momentum, and in the step from (\ref{pro}) to (\ref{whichr}) we have used the definition of the symmetric product or inner product from Geometric Algebra \cite{Clifford}. Comparing the integral in (\ref{52-step}) with the result (\ref{whichr}) in light of the equality (\ref{equabell}), for the case when ${\mathbf L}({\mathbf s}_1)\,{\mathbf L}({\mathbf s}_2)=-1$ is respected, then leads to
\begin{align}
{\cal E}({\mathbf a},{\mathbf b})\,&=\int_{S^2}\!\left\{-\,\text{sign}(\mathbf{a}\cdot\mathbf{s}^i_1)\;\text{sign}(\mathbf{b}\cdot\mathbf{s}^i_2)\right\}\,p({\mathbf s}^i)\;d{\mathbf s}^i \notag \\
&=\,-\cos(\,\eta_{{\mathbf a}{\mathbf b}}). \label{62-con}
\end{align}
Unlike for the weak correlations (\ref{51-con}), for the above correlations the bounds on the Bell-CHSH sum of expectation values are known to exceed the values of $\pm2$ set by Bell's theorem \cite{Bell-1964,Begs}: 
\begin{equation}
-2\sqrt{2}\leqslant{\cal E}({\bf a},\,{\bf b})+{\cal E}({\bf a},\,{\bf b'})+{\cal E}({\bf a'},\,{\bf b})-{\cal E}({\bf a'},\,{\bf b'})\leqslant+2\sqrt{2}\,. \label{44final}
\end{equation}
It is important to bear in mind that this derivation of the singlet correlations is a {\it theoretical} prediction of the 3-sphere model proposed in \cite{Disproof,IJTP,RSOS,IEEE-1,IEEE-2,local}. It should not be confused with after-the-event data analysis followed by experimenters. It is also worth noting that the above is but one of several ways the strong correlations can be derived within $S^3$ \cite{Disproof,IJTP,RSOS,IEEE-1,IEEE-2,local}.

It is evident from the calculations in equations (\ref{43-say}) to (\ref{50-int}) for Bell's local model and equations (\ref{52-step}) to (\ref{whichr}) for the 3-sphere model that in both cases the correct predictions of correlations are obtained from essentially the same set of equations, apart from the condition ${\bf L}({\bf s}_1){\bf L}({\bf s}_2)=-1$ used in the case of the 3-sphere model. However, because of the use of this condition, equations (\ref{54-step}) to (\ref{whichr}) may give the impression that, unlike in the derivation of (\ref{50-int}), the spins ${\bf L}({\bf s}_1)$ and ${\bf L}({\bf s}_2)$ do not play any role in the derivation of (\ref{whichr}). But by setting the value of the product ${\mathbf L}({\mathbf s}_1)\,{\mathbf L}({\mathbf s}_2)$ different from $-1$ so that the conservation of spin angular momentum is violated, we can verify that the spins do play important role in both derivations. To that end, let us set ${\mathbf L}({\mathbf s}_1)\,{\mathbf L}({\mathbf s}_2)=+0.3$ for the sake of argument. Then it is straightforward to verify by following the steps (\ref{52-step}) to (\ref{whichr}) that, in this case, the final result (\ref{whichr}) would be $+0.3\cos(\,\eta_{{\mathbf a}{\mathbf b}})$ instead of $-\cos(\,\eta_{{\mathbf a}{\mathbf b}})$. This demonstrates that the spins ${\bf L}({\bf s}_1)$ and ${\bf L}({\bf s}_2)$ do play a fundamental role in both derivations. For further explanation, see also Appendix~\ref{QandA}.

\section{Comparison of the 3-sphere model with Bell's local model}

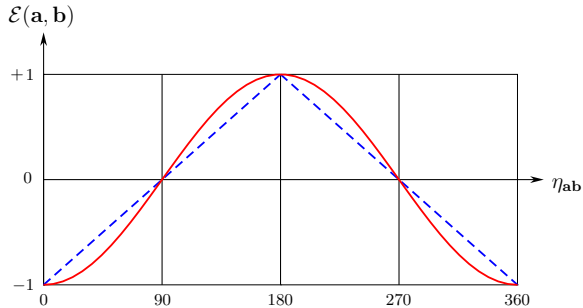
\begin{figure}[t]
\hrule
\scalebox{0.7}{
\begin{pspicture}(0.0,-1.5)(9.6,6.4)
\psset{xunit=0.5mm,yunit=4cm}
\psaxes[axesstyle=frame,linewidth=0.01mm,tickstyle=full,ticksize=0pt,dx=90\psxunit,Dx=180,dy=1
\psyunit,Dy=+2,Oy=-1](0,0)(180,1.0)
\psline[linewidth=0.2mm,arrowinset=0.3,arrowsize=2pt 3,arrowlength=2]{->}(0,0.5)(190,0.5)
\psline[linewidth=0.2mm]{-}(45,0)(45,1)
\psline[linewidth=0.2mm]{-}(90,0)(90,1)
\psline[linewidth=0.2mm]{-}(135,0)(135,1)
\psline[linewidth=0.2mm,arrowinset=0.3,arrowsize=2pt 3,arrowlength=2]{->}(0,0)(0,1.2)
\psline[linewidth=0.35mm,linestyle=dashed,linecolor=blue]{-}(0,0)(90,1)
\psline[linewidth=0.35mm,linestyle=dashed,linecolor=blue]{-}(90,1)(180,0)
\put(2.1,-0.38){${90}$}
\put(6.5,-0.38){${270}$}
\put(-0.63,3.92){${+}$}
\put(-0.67,5.0){\large{${\cal E}{({\bf a},{\bf b})}$}}
\put(-0.38,1.93){${0}$}
\put(9.65,1.85){\large ${\eta_{{\bf a}{\bf b}}}$}
\psplot[linewidth=0.35mm,linecolor=red]{0.0}{180}{x dup cos exch cos mul 1.0 mul neg 1 add}
\end{pspicture}}
\hrule
\caption{Graphs of the correlations (\ref{51-con}) and (\ref{62-con}). The x-axis depicts the angle in degrees between the detector directions ${\mathbf a}$ and ${\mathbf b}$, and the y-axis depicts the corresponding expectation value or correlations. The dotted straight lines depict the correlations predicted by Bell's local model and the solid curve represents the predictions of the 3-sphere model for which ${\mathbf L}({\mathbf s}_1)\,{\mathbf L}({\mathbf s}_2)=-1$ holds during the free evolution of the spins. After \cite{IJTP,IEEE-2}.}
\label{Fig-2}
\hrule
\end{figure}

In the previous section, we derived two contradictory results in equations (\ref{51-con}) and (\ref{62-con}). Remarkably, the left-hand sides of these two equations are mathematically identical. Their right-hand sides, however, differ significantly, and this difference is manifest from their graphs shown in Fig.~\ref{Fig-2}. We also identified the physical and mathematical reason for this difference, namely, the conservation of initial zero spin angular momentum encoded in the condition ${\mathbf L}({\mathbf s}_1)\,{\mathbf L}({\mathbf s}_2)=-1$, which is the only global feature of $S^3$ that plays a direct role in the derivation of the correlations (\ref{62-con}) (see also Appendix~\ref{QandA}).

\begin{figure}
\hrule
\scalebox{1}{
\begin{pspicture}(0.0,0.0)(5.3,5.5)

\psarc[linecolor=green](2.5,2.5){2.0}{-16}{41}

\psarc[linecolor=yellow](2.5,2.5){2.0}{41}{162}

\psarc[linecolor=green](2.5,2.5){2.0}{162}{219}

\psarc[linecolor=yellow](2.5,2.5){2.0}{219}{344}

\psellipticarc[rot=-17,linestyle=dashed,linewidth=0.5pt,linecolor=blue](2.5,2.5)(2,0.6){0}{180}

\psellipticarc[rot=-17,linestyle=solid,linecolor=blue](2.5,2.5)(2,0.6){180}{360}

\psellipticarc[rot=40,linestyle=dashed,linewidth=0.5pt,linecolor=red](2.5,2.5)(2,0.7){0}{180}

\psellipticarc[rot=40,linestyle=solid,linecolor=red](2.5,2.5)(2,0.7){180}{360}

\psline[linewidth=0.3mm,arrowinset=0.2,arrowsize=1pt 3,arrowlength=2,linecolor=red]{->}(2.5,2.5)(0.6,4.5)

\psline[linewidth=0.3mm,arrowinset=0.2,arrowsize=1pt 3,arrowlength=2,linecolor=blue]{->}(2.5,2.49)(3.4,5.1)

\put(1.9,1.7){{${\eta_{{\mathbf a}{\mathbf b}}}$}}

\put(2.03,3.50){{${\eta_{{\mathbf a}{\mathbf b}}}$}}

\put(3.45,4.8){{${\mathbf b}$}}

\put(0.89,4.35){{${\mathbf a}$}}

\put(-1.0,4.3){${{\rm I\!R}^3}$}

\put(4.46,0.48){{${S^2}$}}

\put(1.9,3.95){${\textcolor{red}{+}\,\textcolor{blue}{-}}$}

\put(0.57,2.2){${\textcolor{red}{+}\,\textcolor{blue}{+}}$}

\put(3.85,2.65){${\textcolor{red}{-}\,\textcolor{blue}{-}}$}

\put(2.5,0.9){${\textcolor{red}{-}\,\textcolor{blue}{+}}$}

\psarc[linewidth=0.3mm,arrowinset=0.3,arrowsize=2pt 3,arrowlength=1,linecolor=gray]{<-}(4.3,1.3){0.7}{215}{280}

\psdots[linecolor=gray,dotsize=2pt ](2.5,2.5)

\psellipticarc[rot=40,linewidth=0.3mm,arrowinset=0.3,arrowsize=2pt 3,arrowlength=1,linecolor=yellow]{-}(1.355,3.7)(0.075,0.05){120}{60}

\psellipticarc[rot=-17,linewidth=0.3mm,arrowinset=0.3,arrowsize=2pt 3,arrowlength=1,linecolor=yellow]{-}(3.035,4.05)(0.075,0.05){115}{55}

\psarc[linecolor=gray]{<->}(2.5,2.5){0.8}{70}{134}

\psarc[linecolor=gray]{<->}(2.7,1.84){0.95}{162}{215}

\end{pspicture}}
\hrule
\caption{The figure 6.5 of Peres \cite{Peres}, with appropriate changes in notation. The calculation of various probabilities of obtaining measurement results in Bell \cite{Bell-1964} and Peres \cite{Peres} relies on a unit $S^2$ embedded in ${{\rm I\!R}^3}$. In our quaternionic 3-sphere model hypothesized in \cite{Disproof,IJTP,RSOS,IEEE-1,IEEE-2,local}, the unit $S^2$ of Bell and Peres plays the role of a base manifold in the Hopf bundle of $S^3$. For a given pair $\{{\bf a},\,{\bf b}\}$ of measurement directions, $S^2$ is divided into four sectors, cut by two equatorial planes perpendicular to ${\bf a}$ and ${\bf b}$, with alternating signs for the product ${\mathscr A}({\bf a},\,{\mathbf s}^i_1)\,{\mathscr B}({\bf b},\,{\mathbf s}^i_2)$. After \cite{Peres}.}
\label{Fig-3}
\hrule
\end{figure}
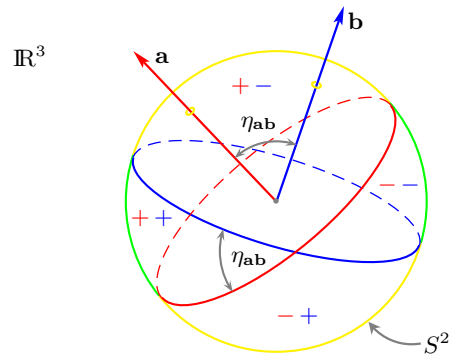

In this section we explain how the constraint ${\mathbf L}({\mathbf s}_1)\,{\mathbf L}({\mathbf s}_2)=-1$ induces the spinorial sign changes (\ref{signchanges}) in quaternions that have been neglected by Bell \cite{Bell-1964} and Peres \cite{Peres} in their derivation (\ref{51-con}) of the correlations within $S^2\hookrightarrow\mathrm{I\!R}^3$. Note that in their $S^2$ model spins are represented inadequately by ordinary vectors such as ${\bf s}_1$ and ${\bf s}_2$ and the conservation of zero spin is stated as ${\bf s}_1+{\bf s}_2=0$ or as ${\bf s}_1=-{\bf s}_2$, whereas in our $S^3$ model spins are represented by the bivectors $-{\mathbf L}({\mathbf s}_1)=-I_3{\bf s}_1$ and ${\mathbf L}({\mathbf s}_2)=I_3{\bf s}_2$ satisfying the conditions (\ref{55}) and (\ref{56}). Consequently, in our derivation (\ref{62-con}), the spinorial sign changes (\ref{signchanges}) have been taken into account automatically as parts of the geometrical features of $S^3$. However, it is instructive to bring out in which step in their derivation of the correlations (\ref{51-con}) Bell \cite{Bell-1964} and Peres \cite{Peres} have ended up neglecting the spinorial sign changes. For this purpose, it is instructive to reflect on figure 6.5 on page 161 of Peres \cite{Peres}, which has been used for calculating the probabilities of various measurement results jointly observed by Alice and Bob. We have reproduced that figure here for convenience as Fig.~\ref{Fig-3}, with appropriate changes in notation.

Now in Bell's local model, the measurement functions of Alice and Bob are the following sign functions, as in (\ref{79-nom}): 
\begin{equation}
{\mathscr A}({\mathbf a},{\mathbf s}^i_1)=+\mu_1 =\text{sign}(\mathbf{a}\cdot\mathbf{s}^i_1)=\pm1 \label{59-pod}
\end{equation}
and
\begin{equation}
{\mathscr B}({\mathbf b},{\mathbf s}^i_2)=-\mu_2 =-\,\text{sign}(\mathbf{s}^i_2\cdot\mathbf{b})=\mp1. \label{59-nod}
\end{equation}
Thus, as we saw in Section~\ref{Sec-3}, if the initial direction ${\bf s}^i$ of the two spins is uncontrollable but describable by isotropic probability distribution $p({\bf s}^i)$, then the probability that the spin of particle 1 observed by Alice will be detected parallel to the vector ${\mathbf{a}}$ (regardless of whether particle 2 is detected) is unambiguously predicted by Bell's local model to be
\begin{equation}
P_1^{+}({\mathbf{a}})=P_1^{-}({\mathbf{a}})=\frac{1}{2}.
\end{equation}
And, likewise, the probability that the spin of particle 2 observed by Bob will be detected parallel to ${\mathbf{b}}$ is given~by
\begin{equation}
P_2^{+}({\mathbf{b}})=P_2^{-}({\mathbf{b}})=\frac{1}{2}.
\end{equation}
To calculate the joint probabilities of observing the two results defined in (\ref{59-pod}) and (\ref{59-nod}) following Bell \cite{Bell-1964} and Peres \cite{Peres}, consider a unit $S^2$ embedded in ${{\rm I\!R}^3}$, cut through by the equatorial planes perpendicular to the detector directions ${\bf a}$ and ${\bf b}$, as shown in Fig.~\ref{Fig-3}. We then have ${\mathscr A}({\mathbf a},{\mathbf s}^i_1)=+1$ if ${\bf s}^i_1$ points through one of the hemispheres, and ${\mathscr A}({\mathbf a},{\mathbf s}^i_1)=-1$ if it points through the other hemisphere. Likewise, a second equatorial plane perpendicular to ${\bf b}$ determines the regions where ${\mathscr B}({\mathbf b},{\mathbf s}^i_2)=\pm1$. The unit $S^2$ is thereby divided by the two equatorial planes into four sectors, with alternating signs for the product ${\mathscr A}({\mathbf a},{\mathbf s}^i_1)\,{\mathscr B}({\mathbf b},{\mathbf s}^i_2)$. The adjacent sectors have their surface areas proportional to $\eta_{{\mathbf{a}}{\mathbf{b}}}$ and $\pi-\eta_{{\mathbf{a}}{\mathbf{b}}}$ for $0\leqslant\eta_{{\mathbf{a}}{\mathbf{b}}}\leqslant\pi$, and proportional to $2\pi-\eta_{{\mathbf{a}}{\mathbf{b}}}$ and $\pi-(2\pi-\eta_{{\mathbf{a}}{\mathbf{b}}})=\eta_{{\mathbf{a}}{\mathbf{b}}}-\pi$ for $\pi\leqslant\eta_{{\mathbf{a}}{\mathbf{b}}}\leqslant2\pi$, with the area of each hemisphere being $2\pi$. As a result, the probabilities of observing the results jointly and simultaneously work out to be
\begin{align}
P_{12}^{++}(\eta_{{\mathbf{a}}{\mathbf{b}}})&=
\begin{rcases}
\begin{dcases}
&\!\!\!\!\!\frac{\eta_{{\mathbf a}{\mathbf b}}}{2\pi}
\;\;\;\;\;\;\;\;\;\;\;\text{if} \;\;\,0 \leqslant \eta_{{\mathbf a}{\mathbf b}} \leqslant \pi \\
&\!\!\!\!\frac{\pi-\eta_{{\mathbf a}{\mathbf b}}}{2\pi}
\;\;\;\;\text{if} \;\;\pi \leqslant \eta_{{\mathbf a}{\mathbf b}} \leqslant 2\pi
\end{dcases}
\end{rcases} \label{3} \\
&=P_{12}^{--}(\eta_{{\mathbf{a}}{\mathbf{b}}}) \notag
\end{align}
with $P_{12}^{++}(\eta_{{\mathbf{a}}{\mathbf{b}}})\equiv P_{12}\{{\mathscr A}=+1,\;{\mathscr B}=+1
\;|\;\eta_{{\mathbf{a}}{\mathbf{b}}}\}$, and
\begin{align}
P_{12}^{+-}(\eta_{{\mathbf{a}}{\mathbf{b}}})&=
\begin{rcases}
\begin{dcases}
&\!\!\!\!\frac{2\pi-\eta_{{\mathbf a}{\mathbf b}}}{2\pi}
\;\;\;\text{if} \;\;\,0 \leqslant \eta_{{\mathbf a}{\mathbf b}} \leqslant \pi \\
&\!\!\!\!\frac{\eta_{{\mathbf a}{\mathbf b}}-\pi}{2\pi}
\;\;\;\;\;\text{if} \;\;\pi \leqslant \eta_{{\mathbf a}{\mathbf b}} \leqslant 2\pi
\end{dcases}
\end{rcases} \label{4} \\
&=P_{12}^{-+}(\eta_{{\mathbf{a}}{\mathbf{b}}}), \notag
\end{align}
with $P_{12}^{+-}(\eta_{{\mathbf{a}}{\mathbf{b}}})\equiv P_{12}\{{\mathscr A}=+1,\;{\mathscr B}=-1
\;|\;\eta_{{\mathbf{a}}{\mathbf{b}}}\}$, where ${P_{12}^{+-}(\eta_{{\mathbf{a}}{\mathbf{b}}})}$, {\it etc.}, are probabilities of observing the result $+1$ by Alice and $-1$ by Bob, {\it etc}., and the subscripts 1 and 2 label the two remote observation stations of Alice and Bob as in Fig.~\ref{Fig-1}. The remaining probabilities all vanish:
\begin{align}
P_{12}^{+0}(\eta_{{\mathbf{a}}{\mathbf{b}}})=P_{12}^{-0}(\eta_{{\mathbf{a}}{\mathbf{b}}})&=P_{12}^{0+}(\eta_{{\mathbf{a}}{\mathbf{b}}})=P_{12}^{0-}(\eta_{{\mathbf{a}}{\mathbf{b}}}) \notag \\
&= P_{12}^{00}(\eta_{{\mathbf{a}}{\mathbf{b}}})=0,
\end{align}
where the superscript 0 stands for no detection. The correlations predicted by Bell's local model thus seem to be
\begin{widetext}
\begin{align}
{\cal E}({\mathbf a},\,{\mathbf b})\!&=\!\!\int
{\mathscr A}({\mathbf a},\,{\mathbf s}^i_1)\,{\mathscr B}({\mathbf b},\,{\mathbf s}^i_2)\;p({\mathbf s}^i)\,d{\mathbf s}^i\,\approx\lim_{\,n\,\gg\,1}\left[\frac{1}{n}\sum_{k\,=\,1}^{n}\,{\mathscr A}({\mathbf a},\,{\mathbf s}_1^k)\;{\mathscr B}({\mathbf b},\,{\mathbf s}_2^k)\right] \\
&=\frac{({\mathscr A}{\mathscr B}=++)\!\times\! P^{++}(\eta_{{\mathbf{a}}{\mathbf{b}}})\!+\!({\mathscr A}{\mathscr B}=--)\!\times\!P^{--}(\eta_{{\mathbf{a}}{\mathbf{b}}})\!+\!({\mathscr A}{\mathscr B}=+-)\!\times\! P^{+-}(\eta_{{\mathbf{a}}{\mathbf{b}}})\!+\!({\mathscr A}{\mathscr B}=-+)\!\times\!P^{-+}(\eta_{{\mathbf{a}}{\mathbf{b}}})}{\left\{P^{++}(\eta_{{\mathbf{a}}{\mathbf{b}}})\,+\,P^{--}(\eta_{{\mathbf{a}}{\mathbf{b}}})\,+\,P^{+-}(\eta_{{\mathbf{a}}{\mathbf{b}}})\,+\,P^{-+}(\eta_{{\mathbf{a}}{\mathbf{b}}})\right\}=1} \\
\!&=
\!\begin{dcases}
\!\left[\left\{(++)\!\times\!\left(\frac{\eta_{{\bf a}{\bf b}}}{2\pi}\right)\!\right\}+\left\{(--)\!\times\!\left(\frac{\eta_{{\bf a}{\bf b}}}{2\pi}\right)\!\right\}
+\left\{(+-)\!\times\!\left(\!\frac{\pi-\eta_{{\bf a}{\bf b}}}{2\pi}\!\right)\!\right\}+\left\{(-+)\!\times\!\left(\!\frac{\pi-\eta_{{\bf a}{\bf b}}}{2\pi}\!\right)\!\right\}\right]& \text{if}\;\,0 \leqslant \eta_{{\mathbf a}{\mathbf b}} \leqslant \pi \label{66-nod} \\
\!\left[\!\left\{\!(++)\!\times\!\!\left(\!\frac{2\pi\!-\!\eta_{{\mathbf a}{\mathbf b}}}{2\pi}\!\right)\!\right\}\!+\!\left\{\!(--)\!\times\!\!\left(\!\frac{2\pi\!-\!\eta_{{\mathbf a}{\mathbf b}}}{2\pi}\!\right)\!\right\}
\!+\!\left\{\!(+-)\!\times\!\!\left(\!\frac{\eta_{{\mathbf a}{\mathbf b}}\!-\!\pi}{2\pi}\!\right)\!\right\}\!+\!\left\{\!(-+)\!\times\!\!\left(\!\frac{\eta_{{\mathbf a}{\mathbf b}}\!-\!\pi}{2\pi}\!\right)\!\right\}\!\right]\!\!\!\!\!& \text{if} \;\pi \leqslant \eta_{{\mathbf a}{\mathbf b}} \leqslant 2\pi
\end{dcases} \\
&=
\begin{dcases}
-\,1\,+\,\frac{2}{\pi}\,\eta_{{\mathbf a}{\mathbf b}}
\;\;\;\text{if} &\!\! 0 \leqslant \eta_{{\mathbf a}{\mathbf b}} \leqslant \pi \\
+\,3\,-\,\frac{2}{\pi}\,\eta_{{\mathbf a}{\mathbf b}}
\;\;\;\text{if} &\!\! \pi \leqslant \eta_{{\mathbf a}{\mathbf b}} \leqslant 2\pi\,,
\end{dcases}
\end{align}
\end{widetext}
$\!\!\!$where ${n}$ is the total number of experiments performed, $k$ is the trial number, $p({\bf s}^i)=\frac{1}{n}$ is the probability density, and $({\mathscr A}{\mathscr B}=+-)$, {\it etc.}, indicate the values of the products such as ${\mathscr A}{\mathscr B}=-1$ for the joint results ${\mathscr A}=+1$ and ${\mathscr B}=-1$, {\it etc}.

As noted, these weak correlations are depicted by dashed straight lines in Fig.~\ref{Fig-2}. In our derivation (\ref{51-con}) within $S^3$, they follow if the conservation of spin angular momentum encoded in the condition ${\mathbf L}({\mathbf s}_1)\,{\mathbf L}({\mathbf s}_2)=-1$ is neglected, which amounts to neglecting the spinorial sign changes (\ref{signchanges}) intrinsic to the quaternions that constitute the 3-sphere. On the other hand, if the spinorial sign changes are taken into account in the derivation by respecting the condition ${\mathbf L}({\mathbf s}_1)\,{\mathbf L}({\mathbf s}_2)=-1$, then the resulting correlations are ${\cal E}({\mathbf a},{\mathbf b})=-\cos(\,\eta_{{\mathbf a}{\mathbf b}})$, as we derived in (\ref{62-con}). They are depicted by the sinusoidal curve in Fig.~\ref{Fig-2}. We can appreciate this in more detail as follows. Using the mathematical identity
\begin{align}
&-\cos\left(\eta_{{\bf a}{\bf b}}\right) \notag \\
&\equiv \frac{\frac{1}{2}\sin^2\!\left(\frac{\eta_{{\bf a}{\bf b}}}{2}\right)\!+\!\frac{1}{2}\sin^2\!\left(\frac{\eta_{{\bf a}{\bf b}}}{2}\right)
\!-\!\frac{1}{2}\cos^2\!\left(\frac{\eta_{{\bf a}{\bf b}}}{2}\right)\!-\!\frac{1}{2}\cos^2\!\left(\frac{\eta_{{\bf a}{\bf b}}}{2}\right)}{\frac{1}{2}\sin^2\!\left(\frac{\eta_{{\bf a}{\bf b}}}{2}\right)
\!+\!\frac{1}{2}\sin^2\!\left(\frac{\eta_{{\bf a}{\bf b}}}{2}\right)
\!+\!\frac{1}{2}\cos^2\!\left(\frac{\eta_{{\bf a}{\bf b}}}{2}\right)
\!+\!\frac{1}{2}\cos^2\!\left(\frac{\eta_{{\bf a}{\bf b}}}{2}\right)},
\end{align}
the correlations we derived in (\ref{62-con}) can be expressed as
\begin{widetext}
\begin{align}
{\cal E}({\mathbf a},\,{\mathbf b})\!&=-\cos(\,\eta_{{\mathbf a}{\mathbf b}}) \\
&=\,\frac{\left[(++)\!\times\!\left\{\frac{1}{2}\sin^2\!\left(\frac{\eta_{{\bf a}{\bf b}}}{2}\right)\right\}\right]+\left[(--)\!\times\!\left\{\frac{1}{2}\sin^2\!\left(\frac{\eta_{{\bf a}{\bf b}}}{2}\right)\right\}\right]
+\left[(+-)\!\times\!\left\{\frac{1}{2}\cos^2\!\left(\frac{\eta_{{\bf a}{\bf b}}}{2}\right)\right\}\right]+\left[(-+)\times\left\{\frac{1}{2}\cos^2\!\left(\frac{\eta_{{\bf a}{\bf b}}}{2}\right)\right\}\right]}{\left[\frac{1}{2}\sin^2\!\left(\frac{\eta_{{\bf a}{\bf b}}}{2}\right)
+\frac{1}{2}\sin^2\!\left(\frac{\eta_{{\bf a}{\bf b}}}{2}\right)
+\frac{1}{2}\cos^2\!\left(\frac{\eta_{{\bf a}{\bf b}}}{2}\right)
+\frac{1}{2}\cos^2\!\left(\frac{\eta_{{\bf a}{\bf b}}}{2}\right)\right]}. \label{69-com}
\end{align}
\end{widetext}
Comparing this expression with (\ref{66-nod}) we thus see that, for $0 \leqslant \eta_{{\mathbf a}{\mathbf b}} \leqslant \pi$, the probability of obtaining the joint result ${\mathscr A}{\mathscr B}=++$ is not $\frac{\eta_{{\bf a}{\bf b}}}{2\pi}$ but $\frac{1}{2}\sin^2\!\left(\frac{\eta_{{\bf a}{\bf b}}}{2}\right)$ if the spinorial sign changes are taken into account. In other words, occasionally what may be deemed to be a result ${\mathscr A}{\mathscr B}=++$ according to the reasoning in Bell \cite{Bell-1964} and Peres \cite{Peres}, would in fact be either ${\mathscr A}{\mathscr B}=+-$ or ${\mathscr A}{\mathscr B}=-+$, thereby changing the number of times the result ${\mathscr A}{\mathscr B}=++$ occurs from $\frac{\eta_{{\bf a}{\bf b}}}{2\pi}$-many times to $\frac{1}{2}\sin^2\!\left(\frac{\eta_{{\bf a}{\bf b}}}{2}\right)$-many times. We can appreciate this by considering the example of the result ${\mathscr A}$ defined in (\ref{79-nmn}):
\begin{align}
S^3\ni{\mathscr A}({\mathbf a},{\mathbf s}^i_1)\,&=\lim_{{\mathbf s}_1\,\rightarrow\,\mu_1{\mathbf a}}\left\{-\,{\mathbf D}({\mathbf a})\,{\mathbf L}({\mathbf s}_1)\right\} \notag \\
&=\lim_{{\mathbf s}_1\,\rightarrow\,\mu_1{\mathbf a}}\left\{\,+\,{\mathbf q}(\eta_{{\mathbf a}{\mathbf s}_1},\,{\mathbf r}_1)\right\}. \label{70-res}
\end{align}
Here the spin ${\mathbf L}({\mathbf s}_1)$ is a bivector or pure quaternion, and corresponds to a binary rotation --- {\it i.e.}, rotation by $\pi$ \cite{Altmann}. We can recognize this by considering the corresponding non-pure quaternion and noticing that ${\bf q}(\beta=\pi,\,{\mathbf s}_1)={\mathbf L}({\mathbf s}_1)$:
\begin{equation}
{\bf q}(\beta,\,{\mathbf s}_1)=\cos\left(\frac{\beta}{2}\right)+{\mathbf L}({\mathbf s}_1)\,\sin\left(\frac{\beta}{2}\right)\!.
\end{equation}
But the product of two rotations by $\pi$ about the same axis ${\bf s}_1$ is a rotation by $2\pi$, and that is {\it not} an identity operation but subject to spinorial sign changes specified in (\ref{signchanges}). Indeed, any product of two unit bivectors, such as ${\mathbf L}({\mathbf s}_1)\,{\mathbf L}({\mathbf s}_2)$ for ${\bf s}_1={\bf s}_2$, is equal to $-1$, which is precisely our condition (\ref{56}) for the conservation of angular momentum. Thus the condition ${\mathbf L}({\mathbf s}_1)\,{\mathbf L}({\mathbf s}_2)=-1$ enforces sign changes in the results (\ref{70-res}) because they depend on ${\mathbf L}({\mathbf s}_1)$, thereby inducing changes also in the products of results such as ${\mathscr A}{\mathscr B}=++\,\longrightarrow\,-+$, {\it etc}. Consequently, the probability of occurring the joint result ${\mathscr A}{\mathscr B}=++$ changes from $\frac{\eta_{{\bf a}{\bf b}}}{2\pi}$ to $\frac{1}{2}\sin^2\!\left(\frac{\eta_{{\bf a}{\bf b}}}{2}\right)$, as we noted by comparing the expressions (\ref{66-nod}) and (\ref{69-com}).

Now a sign change in a bivector or pure quaternion, such as ${\mathbf L}({\mathbf s}_1)$, means a change in its sense of rotation. In other words, a counterclockwise rotation, say ${+\mathbf L}({\mathbf s}_1)$, changes to a clockwise rotation, ${-\mathbf L}({\mathbf s}_1)$. This can be expressed as
\begin{equation}
+\,{\bf L}({\bf s}_1)=+I_3{\bf s}_1 \;\longrightarrow\; -\,{\bf L}({\bf s}_1)=-\,I_3{\bf s}_1=I_3(-\,{\bf s}_1).
\end{equation}
Thus a spinorial sign change in the bivector $+{\bf L}({\bf s}_1)$ can be expressed equivalently as a sign change in the direction of its axis vector ${\bf s}_1$. Indeed, a counterclockwise rotation about $+\,{\bf s}_1$ represented by $+{\bf L}({\bf s}_1)$ is the same as the clockwise rotation about $-\,{\bf s}_1$. But that implies that sign changes in the local result such as (\ref{70-res}) will induce changes such as 
\begin{equation}
+\,\text{sign}(\mathbf{a}\cdot\mathbf{s}^i_1) \;\longrightarrow\; -\,\text{sign}(\mathbf{a}\cdot\mathbf{s}^i_1)
\end{equation}
in the integrand on the left-hand side of (\ref{51-con}), so that some of its values will change from $++$ to $+-$, {\it etc.}, under the spinorial sign changes seen in (\ref{signchanges}), with precisely which changes would occur dictated by the condition ${\mathbf L}({\mathbf s}_1)\,{\mathbf L}({\mathbf s}_2)=-1$ for the conservation of spin angular momentum we imposed in the step (\ref{54-step}) of our derivation of the correlations (\ref{62-con}).

\section{Recovering Bell's Local Model from the $S^3$ Model in the Flat Geometry of $\mathrm{I\!R}^3$} \label{Hopf}

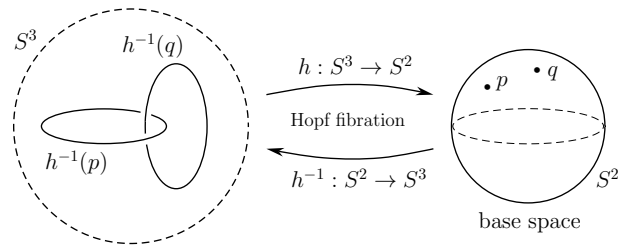
\begin{figure}
\hrule
\scalebox{0.6}{
\begin{pspicture}(0.0,-4.0)(4.5,3.3)

\pscircle[linewidth=0.3mm,linestyle=dashed](-1.8,-0.45){2.6}

\psellipse[linewidth=0.3mm](-0.8,-0.45)(0.7,1.4)

\psellipse[linewidth=0.3mm,border=3pt](-2.4,-0.45)(1.4,0.4)

\pscurve[linewidth=0.3mm,border=3pt](-1.485,-0.35)(-1.48,-0.25)(-1.45,0.0)

\pscircle[linewidth=0.3mm](7.0,-0.45){1.7}

\psellipse[linewidth=0.2mm,linestyle=dashed](7.0,-0.45)(1.68,0.4)

\put(-4.4,1.27){{\Large ${S^3}$}}

\put(-2.0,1.2){{\Large ${h^{-1}(q)}$}}

\put(-3.7,-1.4){{\Large ${h^{-1}(p)}$}}

\put(7.43,0.67){{\Large ${q}$}}

\psdot*(7.2,0.79)

\put(6.3,0.43){{\Large ${p}$}}

\psdot*(6.1,0.43)

\put(8.5,-1.8){{\Large ${S^2}$}}

\put(5.9,-2.7){\Large base space}

\put(1.9,0.7){\Large ${h:S^3\rightarrow S^2}$}

\pscurve[linewidth=0.3mm,arrowinset=0.3,arrowsize=3pt 3,arrowlength=2]{->}(1.2,0.25)(2.47,0.45)(3.74,0.45)(4.9,0.25)

\put(1.73,-0.45){\large Hopf fibration}

\pscurve[linewidth=0.3mm,arrowinset=0.3,arrowsize=3pt 3,arrowlength=2]{->}(4.9,-0.95)(3.74,-1.15)(2.47,-1.15)(1.2,-0.95)

\put(1.75,-1.8){\Large ${h^{-1}:S^2\rightarrow S^3}$}

\end{pspicture}}
\hrule
\caption{The tangled web of linked Hopf circles depicting the geometrical and topological non-trivialities of $S^3$. Locally, in the topological sense, ${S^3}$ is isomorphic to the product ${S^2\times S^1}$. Thus ${S^3}$ is ${S^2}$ worth of circles. Each circle ${S^1}$, as fiber $h^{-1}(p)$, threads through every other circle in the bundle ${S^3}$ without sharing a single point with any other circle, and projects down to the point $p$ on ${S^2}$ via Hopf map ${h: S^3\rightarrow S^2}$. Adapted from Ref.~[10].}
\label{fig1}
\hrule
\end{figure}

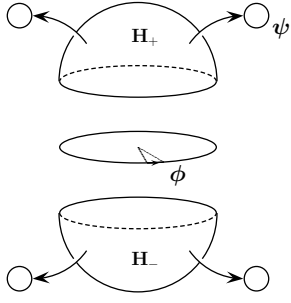
\begin{figure}
\hrule
\scalebox{0.35}{
\begin{pspicture}(3.0,-1.2)(9.0,13.3)

\psellipse[linewidth=0.5mm,border=3pt](6.0,8.5)(3.0,0.625)

\psellipse[linewidth=0.5mm,border=3pt](6.0,6.0)(3.0,0.625)

\psellipse[linewidth=0.5mm,border=3pt](6.0,3.5)(3.0,0.625)

\psarc[linewidth=0.5mm,fillstyle=solid](6.0,8.5){3.0}{0}{180}

\psarc[linewidth=0.5mm,fillstyle=solid](6.0,3.5){3.0}{180}{0}

\psellipticarc[linewidth=0.5mm,linestyle=dashed](6.0,8.5)(3.0,0.625){0}{180}

\psellipticarc[linewidth=0.5mm,linestyle=dashed](6.0,3.5)(3.0,0.625){180}{0}

\psellipticarc[showpoints=true,linewidth=0.5mm](6.0,6.0)(3.0,0.625){300}{330}

\psellipticarc[showpoints=false,arrowscale=2,linewidth=0.3mm]{->}(6.0,6.0)(3.0,0.625){300}{324}

\pscircle[linewidth=0.5mm](1.5,1.0){0.5}

\pscircle[linewidth=0.5mm](10.5,1.0){0.5}

\pscircle[linewidth=0.5mm](1.5,11.0){0.5}

\pscircle[linewidth=0.5mm](10.5,11.0){0.5}

\psarc[linewidth=0.5mm,fillstyle=solid,arrowscale=3]{<-}(10.0,8.5){2.5}{90}{145}

\psarc[linewidth=0.5mm,fillstyle=solid,arrowscale=3]{->}(2.0,8.5){2.5}{35}{90}

\psarc[linewidth=0.5mm,fillstyle=solid,arrowscale=3]{->}(10.0,3.5){2.5}{215}{270}

\psarc[linewidth=0.5mm,fillstyle=solid,arrowscale=3]{<-}(2.0,3.5){2.5}{270}{325}

\put(5.75,10.0){\huge ${\mathbf{H_+}}$}

\put(5.75,1.5){\huge ${\mathbf{H_-}}$}

\put(7.2,4.7){\Huge ${\boldsymbol{\phi}}$}

\put(11.1,10.3){\Huge ${\boldsymbol{\psi}}$}

\end{pspicture}}
\hrule
\caption{\!The Hopf bundle of ${S^3}$ showing two hemispherical neighborhoods ${{\mathrm H}_{\pm}}$ of the base manifold ${S^2}$. A fiber ${S^1=h^{-1}(p)}$ parametrized by angle ${\psi}$ is attached to each point $p$ of ${{\mathrm H}_{\pm}}$. The intersection of ${{\mathrm H}_{\pm}}$ at ${\theta\sim\frac{\pi}{2}}$ is circular a strip parametrized by angle ${\phi\in[0,\,2\pi)}$. Adapted from Ref.~[10].}
\label{fig2}
\hrule
\end{figure}

In addition to what has been discussed above, there is a further, purely geometrical means to recover the traditional interpretation of Bell's theorem within the flat geometry of $\mathrm{I\!R}^3$ from the quaternionic 3-sphere model, by recognizing that the strong singlet correlations are a direct consequence of the M\"obius-like twists in the Hopf bundle of $S^3$ [10]. To appreciate it, recall that, geometrically, $S^3$ is an intricately interlinked bundle of circles, $S^1$, called Hopf circles or Clifford parallels, which are the fibers of the Hopf bundle, with $S^2$ being the base of this bundle, so that locally (in the topological sense) $S^3$ is isomorphic to the product space $S^2\times{S^1}$ (see Fig.~\ref{fig1}). Globally, however, $S^3$ has no cross-section at all. Consequently, each Hopf circle $S^1$ threads through every other circle in the bundle without sharing a single point with any of them, thereby producing M\"obius-like twists in the bundle. These twists, and therefore the Hopf bundle, can be constructed using the relation 
\begin{equation}
e^{i\psi_-}=\,e^{i\omega\phi}\,e^{i\psi_+}\,, \label{A4}
\end{equation}
as shown in Fig.~\ref{fig2}, where ${\psi_-}$ and ${\psi_+}$ in the range $0\leqslant\psi<4\pi$ are the coordinates of the fibers $S^1$ attached to the hemispheres ${{\mathrm H}_-}$ and ${{\mathrm H}_+}$ of the base space ${S^2}$, which is parameterized by spherical coordinates ${(0\leqslant\theta < \pi,\;0\leqslant\phi < 2\pi)}$; ${\phi}$ is the angle that parameterizes a thin intersection ${{\mathrm H}_-\cap {\mathrm H}_+}$ around the equator of ${S^2}$ [${\theta\sim\frac{\pi}{2}}$]; and ${e^{i\omega\phi}}$ is the transition function that glues the two sections ${{\mathrm H}_-}$ and ${{\mathrm H}_+}$ together, thus constituting the 3-sphere. The winding number $\omega$ must be an integer for $\psi$ to be single-valued and the resulting structure to be a manifold. It is evident from the relation (\ref{A4}) that, for $\omega=1$, the fibers $S^1$ match perfectly at the angle ${\phi=0}$ (modulo ${2\pi}$), but differ from each other at all intermediate angles ${\phi}$. For example, ${e^{i\psi_-}}$ and ${e^{i\psi_+}}$ differ by a minus sign at ${\phi=\pi}$. On the other hand, for $\omega=0$ the $S^3$ bundle reduces to a trivial bundle: $S^2\times{S^1}$. While geometrically  $S^3$ is a highly non-trivial manifold, algebraically the twists within it can be characterized by the simple condition
\begin{equation}
{\bf L}({\bf s}_1)\,{\bf L}({\bf s}_2)=-1, \label{A5}
\end{equation}
which is the condition (14) for the conservation of spin angular momentum discussed in the paper. I have proved this in two different ways. In Section~VIII of Ref.~[9], I have derived the conservation condition (\ref{A5}) from the relation (\ref{A4}) of the Hopf bundle, and in Appendix~X of Ref.~[10] I have conversely proved that the relation (\ref{A4}) of Hopf bundle stems from the condition (\ref{A5}) for the conservation of vanishing spin angular momentum.

Now, as we saw in the paper, Bell's local model is set within a 2-sphere, $S^2$, embedded in $\mathrm{I\!R}^3$. In the $S^3$ model, on the other hand, the points of this $S^2$ are projections of the intricately interlinked fibers of $S^3$ by the Hopf map $h$, as shown in Fig.~\ref{fig1}:
\begin{equation}
\mathrm{I\!R}^4\hookleftarrow{S^3}\xrightarrow{\;\;h\;\;}S^2\hookrightarrow{\mathrm{I\!R}}^3.
\end{equation}
If we parameterize $S^3$ with the coordinates $q_1$, $q_2$, $q_3$, and $q_4$ of $\mathrm{I\!R}^4$ so that $q_1^2+q_2^2+q_3^2+q_4^2=1$, then $h\!:S^3\!\rightarrow{S^2}$ is defined~by
\begin{equation}
h(q_1,\,q_2,\,q_3,\,q_4)=(x,\,y,\,z)\,, \label{A77}
\end{equation}
where (as in the original formula discovered by Hopf in 1931)
\begin{align}
x&=2(q_1q_3+q_2q_4)\,, \\
y&=2(q_2q_3-q_1q_4)\,, \\
\text{and}\;\;\,z&=q_1^2+q_2^2-q_3^2-q_4^2
\end{align}
are the corresponding coordinates in $\mathrm{I\!R}^3$ that parameterize the base space $S^2$ of the Hopf bundle as a unit 2-sphere, and satisfy
\begin{equation}
x^2+y^2+z^2=\left(q_1^2+q_2^2+q_3^2+q_4^2\right)^2=1.
\end{equation}

The Hopf map (\ref{A77}) is thus a surjective map that maps all of the points of a unit $S^3$ embedded in $\mathrm{I\!R}^4$ to the points of a unit $S^2$ embedded in $\mathrm{I\!R}^3$, thereby reducing $S^3$ to $S^2$. In particular, the entire set of points of a Hopf circle $S^1$ in $S^3$ is mapped to a point, say $p$, of the base $S^2$ of the bundle, making $S^1=h^{-1}(p)$ a preimage of that point $p$, as shown in Fig.~\ref{fig1}. As a result, the $S^3$ model can be naturally projected by the Hopf map to Bell's local model set within the flat geometry of $\mathrm{I\!R}^3$. In particular, the measurement results ${\mathscr A}({\mathbf a},{\mathbf s}^i_1)$, defined in the $S^3$ model as
\begin{align}
S^3\ni{\mathscr A}({\mathbf a},{\mathbf s}^i_1)\,&=\lim_{{\mathbf s}_1\,\rightarrow\,\mu_1{\mathbf a}}\left\{\cos(\eta_{{\mathbf a}{\mathbf s}_1})+(I_3{\mathbf r}_1)\sin(\eta_{{\mathbf a}{\mathbf s}_1})\right\} \label{A7} \\
&\longrightarrow\mu_1 =\text{sign}(\mathbf{a}\cdot\mathbf{s}^i_1)\in S^2\hookrightarrow\mathrm{I\!R}^3,
\end{align}
are a part of this projection. We can see this by identifying the components of the quaternion in (\ref{A7}), such as $\cos(\eta_{{\mathbf a}{\mathbf s}_1})=q_1$, {\it etc.}, and expressing them in terms of the Euler angles $(\psi,\theta,\phi)$ using the standard representation of arbitrary rigid rotations:
\begin{align}
q_1&=\cos\left(\frac{\psi+\phi}{2}\right)\cos\frac{\theta}{2}\,, &q_2=\sin\left(\frac{\psi+\phi}{2}\right)\cos\frac{\theta}{2}\,,\\
q_3&=\cos\left(\frac{\psi-\phi}{2}\right)\sin\frac{\theta}{2}\,,
&q_4=\sin\left(\frac{\psi-\phi}{2}\right)\sin\frac{\theta}{2}\,. 
\end{align}
The Hopf map (\ref{A77}) then projects them to the base space $S^2$:
\begin{equation}
x=\cos\phi\sin\theta,\;\;\;\;\;y=\sin\phi\sin\theta,\;\;\;\;\;z=\cos\theta. \label{A9}
\end{equation}
These are the Cartesian coordinates in $\mathrm{I\!R}^3$ of the point $(\theta,\phi)$ on a unit $S^2$ embedded in $\mathrm{I\!R}^3$. Thus, the Euler angle $\psi$ in this representation can be identified with the coordinate of a point in $S^1$, which is projected onto the point $(\theta,\phi)$ of the base $S^2$. In the $S^3$ model, this point is then assigned observable values, $\mu_1 =\text{sign}(\mathbf{a}\cdot\mathbf{s}^i_1)=\pm1$, by the limit ${{\mathbf s}_1\,\rightarrow\,\mu_1{\mathbf a}}$ defined in (\ref{A7}), precisely as assumed in Bell's local model set within $S^2\hookrightarrow\mathrm{I\!R}^3$.

The Hopf map $h:S^3\longrightarrow{S^2}$ can also be expressed as
\begin{equation}
{\mathbf q}(\eta_{{\mathbf a}{\mathbf s}_1},\,{\mathbf r}_1) \longmapsto{\mathbf q}(\eta_{{\mathbf a}{\mathbf s}_1},\,{\mathbf r}_1)\,(I_3{\bf e}_z)\;{\mathbf q}^{\dagger}(\eta_{{\mathbf a}{\mathbf s}_1},\,{\mathbf r}_1)
\end{equation}
(with $\dagger$ indicating conjugation) in terms of the quaternion
\begin{equation}
{\mathbf q}(\eta_{{\mathbf a}{\mathbf s}_1},\,{\mathbf r}_1)=
\cos(\eta_{{\mathbf a}{\mathbf s}_1})+(I_3{\mathbf r}_{1})\sin(\eta_{{\mathbf a}{\mathbf s}_1}) \label{quat97}
\end{equation}
appearing in (\ref{A7}), with respect to the fixed reference bivector $I_3{\bf e}_z$ on $S^2$. The map then projects ${\mathbf q}(\eta_{{\mathbf a}{\mathbf s}_1},\,{\mathbf r}_1)$ to the bivector
\begin{equation}
I_3{\bf v}=\,v_x\;{{\bf e}_y}\,\wedge\,{{\bf e}_z}
\,+\,v_y\;{{\bf e}_z}\,\wedge\,{{\bf e}_x}
\,+\,v_z\;{{\bf e}_x}\,\wedge\,{{\bf e}_y}
\label{mu}
\end{equation}
representing a point on $S^2$, whose coordinates in $\mathrm{I\!R}^3$ are
\begin{align}
v_x&=2\,r_{1x}\,r_{1z}\sin^2(\eta_{{\mathbf a}{\mathbf s}_1}\!) - 2\,r_{1y}\cos(\eta_{{\mathbf a}{\mathbf s}_1}\!)\sin(\eta_{{\mathbf a}{\mathbf s}_1}), \\
v_y&=2\,r_{1x}\cos(\eta_{{\mathbf a}{\mathbf s}_1}\!)\sin(\eta_{{\mathbf a}{\mathbf s}_1}\!)+2\,r_{1y}\,r_{1z}\sin^2(\eta_{{\mathbf a}{\mathbf s}_1}), \\
v_z&=\cos^2(\eta_{{\mathbf a}{\mathbf s}_1}) + \left(r_{1z}^2-\,r_{1x}^2-\,r_{1y}^2\right)\sin^2(\eta_{{\mathbf a}{\mathbf s}_1}).
\end{align}

This Hopf projection from $S^3$ to $S^2$, however, does not quite recover Bell's model, because the condition ${\bf L}({\bf s}_1)\,{\bf L}({\bf s}_2)=-1$ is still satisfied, which induces the M\"obius-like twists in the fibers of $S^3$ and thus is responsible for the strong singlet correlations, as we saw in their derivation in equations (57) to (63). But there are no such twists in the flat geometry of $\mathrm{I\!R}^3$, which is taken for granted in Bell's local model. Therefore, we must relinquish the condition ${\bf L}({\bf s}_1)\,{\bf L}({\bf s}_2)=-1$, which is equivalent to setting the winding number ${\omega}=0$ in (\ref{A4}) so that $e^{i\psi_-}\!=e^{i\psi_+}$ for all azimuthal angles $\phi$, to recover Bell's model from the $S^3$ model, thus reducing $S^3$ globally to the trivial product space $S^2\times{S^1}$:
\begin{equation}
S^3\xrightarrow[\equiv\,\omega\,=\,0]{\;\;\;{\bf L}({\bf s}_1)\,{\bf L}({\bf s}_2)\,\not=\,-1\;\;\;}\,S^2 \times S^1.
\end{equation}
The resulting bipartite correlations are then no longer strong,
\begin{align}
-\cos(\,\eta_{{\mathbf a}{\mathbf b}})\;\longrightarrow\;
\begin{cases}
-\,1\,+\,\frac{2}{\pi}\,\eta_{{\mathbf a}{\mathbf b}}
\;\;\;\text{if} &\!\! 0 \leqslant \eta_{{\mathbf a}{\mathbf b}} \leqslant \pi, \\
+\,3\,-\,\frac{2}{\pi}\,\eta_{{\mathbf a}{\mathbf b}}
\;\;\;\text{if} &\!\! \pi \leqslant \eta_{{\mathbf a}{\mathbf b}} \leqslant 2\pi,
\end{cases}
\end{align}
as we proved via their derivation in equations (49) to (56). This completes the recovery of Bell's local model from the $S^3$ model:
\begin{equation}
{\mathrm{I\!R}}^4\hookleftarrow{S^3}\xrightarrow[\equiv\,\omega\,=\,0]{\;\;\;{\bf L}({\bf s}_1)\,{\bf L}({\bf s}_2)\,\not=\,-1\;\;\;}\,S^2\times{S^1}\!\xrightarrow{\;\;h\;\;}\, S^2\hookrightarrow{\mathrm{I\!R}}^3.
\end{equation}
We thus see that the $S^3$ model not only reproduces Bell's local model in the flat geometry of $\mathrm{I\!R}^3$, but also extends and enriches it to reproduce the strong correlations predicted by quantum mechanics if the M\"obius-like twists in $S^3$ are taken into account.

\section{Proposed Macroscopic Test of the 3-sphere Hypothesis}

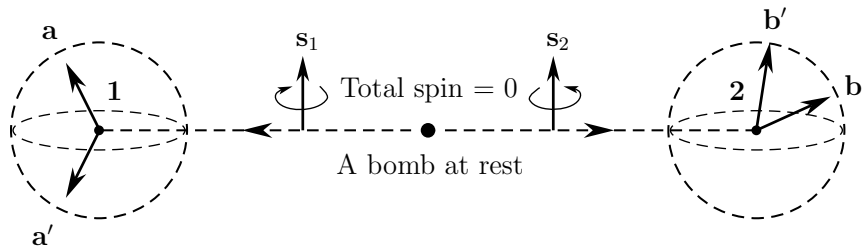
\begin{figure*}
\hrule
\scalebox{0.9}{
\begin{pspicture}(0.0,-2.5)(5.0,2.5)

\psline[linewidth=0.1mm,dotsize=3pt 4]{*-}(-2.51,0)(-2.5,0)

\psline[linewidth=0.1mm,dotsize=3pt 4]{*-}(7.2,0)(7.15,0)

\psline[linewidth=0.4mm,arrowinset=0.3,arrowsize=3pt 3,arrowlength=2]{->}(-2.5,0)(-3,1)

\psline[linewidth=0.4mm,arrowinset=0.3,arrowsize=3pt 3,arrowlength=2]{->}(-2.5,0)(-3,-1)

\psline[linewidth=0.4mm,arrowinset=0.3,arrowsize=3pt 3,arrowlength=2]{->}(7.2,0)(8.3,0.5)

\psline[linewidth=0.4mm,arrowinset=0.3,arrowsize=3pt 3,arrowlength=2]{->}(7.2,0)(7.4,1.3)

\psline[linewidth=0.4mm,arrowinset=0.3,arrowsize=2pt 3,arrowlength=2]{->}(4.2,0)(4.2,1.1)

\psline[linewidth=0.4mm,arrowinset=0.3,arrowsize=2pt 3,arrowlength=2]{->}(0.5,0)(0.5,1.1)

\pscurve[linewidth=0.2mm,arrowinset=0.2,arrowsize=2pt 2,arrowlength=2]{->}(4.0,0.63)(3.85,0.45)(4.6,0.5)(4.35,0.65)

\put(4.1,1.25){{\large ${{\bf s}_2}$}}

\pscurve[linewidth=0.2mm,arrowinset=0.2,arrowsize=2pt 2,arrowlength=2]{<-}(0.35,0.65)(0.1,0.47)(0.86,0.47)(0.75,0.65)

\put(0.4,1.25){{\large ${{\bf s}_1}$}}

\put(-2.4,+0.45){{\large ${\bf 1}$}}

\put(6.8,+0.45){{\large ${\bf 2}$}}

\put(-3.35,1.35){{\large ${\bf a}$}}

\put(-3.5,-1.7){{\large ${\bf a'}$}}

\put(8.5,0.52){{\large ${\bf b}$}}

\put(7.3,1.5){{\large ${\bf b'}$}}

\psline[linewidth=0.3mm,linestyle=dashed](-2.47,0)(2.1,0)

\psline[linewidth=0.4mm,arrowinset=0.3,arrowsize=3pt 3,arrowlength=2]{->}(-0.3,0)(-0.4,0)

\psline[linewidth=0.3mm,linestyle=dashed](2.6,0)(7.2,0)

\psline[linewidth=0.4mm,arrowinset=0.3,arrowsize=3pt 3,arrowlength=2]{->}(5.0,0)(5.1,0)

\psline[linewidth=0.1mm,dotsize=5pt 4]{*-}(2.35,0)(2.4,0)

\pscircle[linewidth=0.3mm,linestyle=dashed](7.2,0){1.3}

\psellipse[linewidth=0.2mm,linestyle=dashed](7.2,0)(1.28,0.3)

\pscircle[linewidth=0.3mm,linestyle=dashed](-2.51,0){1.3}

\psellipse[linewidth=0.2mm,linestyle=dashed](-2.51,0)(1.28,0.3)

\put(1.0,-0.65){\large A bomb at rest}

\put(1.05,0.5){\large Total spin = 0}

\end{pspicture}}
\hrule
\caption{A bomb, initially at rest, explodes into two unequal fragments carrying opposite spin angular momenta. Measurements of the spin components on each fragment are performed at remote stations ${\bf 1}$ and ${\bf 2}$, along two post-selected directions ${\bf a}$ and ${\bf b}$. The two spins represented by the bivectors $-{\bf L}({\bf s}_1)$ and ${\bf L}({\bf s}_2)$ are shown rotating in opposite senses about the same vector ${\bf s}_1={\bf s}_2$, thanks to the conservation of the initial zero angular momentum. This description using a coordinate system of {\it fixed} handedness specified in the bivector subalgebra (\ref{bi-1-m}) is equivalent to Peres's description in \cite{Peres} with ${\bf s}_2=-{\bf s}_1$, because the bivector $-{\bf L}({\bf s}_1)$ can also be written as ${\bf L}(-{\bf s}_1)$. See also the discussion in footnotes~\ref{foot-2} and \ref{foot-3}, and in Section~\ref{Hopf} and Appendix~\ref{QandA}.}
\smallskip
\hrule
\label{Fig-4}
\end{figure*}

In \cite{IJTP} we proposed a macroscopic experiment that may be able to detect the signatures of the above spinorial sign changes under ${2\pi}$ rotations in the form of strong singlet correlations derived in the previous sections. It is worthwhile to discuss this experiment again in light of the above analysis, with some improved considerations. 

If realized, the experiment would determine whether Bell inequalities are violated for the manifestly local and realistic 3-sphere model we considered above. Needless to say, the proposed experiment has the potential to transform our understanding of the relationship between classical and quantum physics. It is based on a macroscopic variant of the local model considered by Bell \cite{Bell-1964} and Peres \cite{Peres} we discussed above. In our proposal, we will closely follow the example of an exploding bomb discussed by Peres \cite{Peres} (which is essentially a pedagogical illustration of Bell's local model \cite{Bell-1964}). However, our proposal differs from Peres's illustration in one important respect. It involves measurements of the actual spin angular momenta of two fragments of an exploding bomb rather than their normalized spin values ${\pm1}$. The latter are to be computed only after all runs of the experiment are completed, which can be executed either in outer space or in a terrestrial laboratory. In the latter case, the effects of the external gravity of Earth and air resistance would complicate matters, but it may be possible to choose experimental parameters judiciously enough to compensate for such unwanted effects.

With this assumption, consider a ``bomb'' made out of a hollow toy ball of diameter, say, three centimeters. The thin hemispherical shells of uniform density that make up the ball are snapped together at their rims in such a manner that a slight increase in temperature would pop the ball open into its two constituents with considerable force \cite{IJTP}. A small lump of density much greater than the density of the ball is attached on the inner surface of each shell at a random location, so that, when the ball pops open, not only would the two shells propagate with equal and opposite linear momenta orthogonal to their common plane, but would also rotate with equal and opposite spin momenta about a random axis in space, as shown in Fig.~\ref{Fig-4}. The volume of the attached lumps can be as small as a cubic millimeter, whereas their mass can be comparable to the mass of the ball. This will facilitate some ${10^6}$ possible spin directions for the two shells, whose outer surfaces can be decorated with colors to make their rotations easily detectable \cite{IJTP}.

Now consider a large ensemble of such balls, identical in every respect except for the relative locations of the two lumps (affixed randomly on the inner surface of each shell). The balls are then placed over a heater---one at a time---at the center of the experimental setup \cite{Peres}, with the common plane of their shells held perpendicular to the horizontal direction of the setup. Although initially at rest, a slight increase in temperature of each ball will eventually eject its two shells towards the observation stations, situated at a chosen distance in mutually opposite directions. Instead of selecting the directions ${\bf a}$ and ${\bf b}$ for observing spin components, however, one or more contact-less rotational motion sensors---capable of determining the precise direction of rotation---are placed near each of the two stations, interfaced with a computer (an improved setup and further practical details are discussed below). These sensors will determine the exact direction of the spin angular momentum ${{\bf s}^k}$ (or ${-\,{\bf s}^k}$) for each shell in a given explosion, without disturbing them otherwise so that their total angular momentum would remain zero, at a designated distance from the center. The interfaced computers can then record this data, in the form of a 3D map of all such directions, at each station.

Once the actual directions of the angular momenta for a large ensemble of shells on both sides are fully recorded, the two computers are instructed to choose a pair of reference directions, ${\bf a}$ for one station and ${\bf b}$ for the other station---from the two 3D maps of already existing data---and then calculate the corresponding pair of numbers ${-\,\text{sign}({\bf a}\cdot{\bf s}^k_1)=\pm\,1}$ and ${+\,\text{sign}({\bf b}\cdot{\bf s}^k_2)=\pm\,1}$. The standard correlation function for the two bomb fragments with $p({\bf s}^i)=\frac{1}{n}$ can then be calculated as
\begin{align}
{\cal E}({\bf a},\,{\bf b})\,&=\int_{S^2}\left\{-\,\text{sign}(\mathbf{a}\cdot\mathbf{s}^i_1)\;\text{sign}(\mathbf{b}\cdot\mathbf{s}^i_2)\right\}\;p({\mathbf s}^i)\,d{\mathbf s}^i\notag \\
&\approx\lim_{\,n\,\gg\,1}\!\left[\frac{1}{n}\!\sum_{k\,=\,1}^{n}
\left\{-\,\text{sign}({\bf a}\cdot{\bf s}^k_1)\;
\text{sign}({\bf b}\cdot{\bf s}^k_2)\right\}\right]\!, \label{74-nav}
\end{align}
together with
\begin{equation}
{\cal E}({\bf a})=\!\lim_{\,n\,\gg\,1}\!\left[\frac{1}{n}\!\sum_{k\,=\,1}^{n}\{\text{sign}(\,+\,{\bf s}^k_1\cdot{\bf a})\}\right]\,=\,0
\end{equation}
and
\begin{equation}
{\cal E}({\bf b})=\!\lim_{\,n\,\gg\,1}\!\left[\frac{1}{n}\!\sum_{k\,=\,1}^{n}\{\text{sign}(\,-\,{\bf s}^k_2\cdot{\bf b})\}\right]\!=\,0\,,
\end{equation}
where ${n}$ is the total number of experiments performed and $k$ specifies a trial number. As we discussed in the previous sections, na\"ive computation of (\ref{74-nav}) by Bell \cite{Bell-1964} and Peres \cite{Peres} gives the weak correlations (\ref{51-con}), whereas the quaternionic 3-sphere model predicts strong correlations (\ref{62-con}) by taking the spinorial properties of the 3-sphere (\ref{nonsin}) into account.

Let us now turn to the practical problem of determining the direction of rotation of a bomb fragment. In order to minimize the contributions of precession and nutation about the rotation axis of the fragment, the bomb may be composed of two flexible squashy balls instead of a single ball. The two balls can then be squeezed together at the start of a run and released as if they were two parts of the same bomb. This will retain the spherical symmetry of the two constituent balls after the explosion, reducing their precession and nutation effects considerably. Consequently, we assume that during the narrow time window of the detection process the contributions of precession and nutation are negligible. In other words, during this narrow time window the individual spins will remain confined to the plane perpendicular to the horizontal direction of the setup. This is because we will then have ${{\bf s} = {\bf r}\times{\bf p}}$, with ${\bf r}$ specifying the location of the massive lump in the constituent ball and ${\bf p}$ being the ball's linear momentum. We can now exploit these physical constraints to determine the direction of rotation of a constituent ball unambiguously, as follows.

Since only the directions of rotation are relevant for computing the correlation function (\ref{1}), it would be sufficient for our purposes to determine only the direction of the vector ${\bf s}$ at each end of the setup. This can be accomplished by arranging three (or more) successive laser screens perpendicular to the horizontal path of the constituent balls, say about half a centimeter apart, and a few judiciously situated cameras around them. To facilitate the detection of the rotation of a ball as it passes through the screens, the surface of the balls can be decorated with distinctive marks, such as dots of different sizes and colors. Then, when a ball passes through the screens, the entry points of a specific mark on the ball can be recorded by the system of cameras. Since the ball would be spinning while passing through the screens, the entry points of the same mark on the successive screens would be located at different relative positions on the screens. The rotation axis of the ball can therefore be determined unambiguously by determining the plane spanned by the entry points and the right-hand rule. In other words, the rotation axis can be determined as the orthogonal direction to the plane spanned by the entry points, with the sense of rotation determined by the right-hand rule. This procedure of determining the direction of rotation can be followed through manually, or it can be automated with the help of computer software. Finally, the horizontal distance from the center of the setup to the location of the middle of the screens can be taken as the distance of the rotation axis from the center of the setup. This distance would help in establishing the simultaneity of the spin measurements at the two ends of the setup.

Undoubtedly, there would be many sources of errors in a mechanical experiment such as this. But if it is performed carefully enough, then our discussion above strongly suggests that it will refute the prediction (\ref{51-con}), which is based on an incorrect calculation by Bell \cite{Bell-1964} and Peres \cite{Peres}, and vindicate the prediction (\ref{62-con}) derived within the 3-sphere model.

Considering the practical difficulties in performing the above experiment, it may seem that its simulation using computer software might provide easier means to verify the hypothesis~\ref{hyp1}. However, any attempt to simulate the experiment would require modeling physical space as a quaternionic 3-sphere, because, according to the hypothesis, the above experiment is supposed to take place within a non-flat $S^3$, not within a flat $\mathrm{I\!R}^3$. Providing the non-commutative background geometry of $S^3$ would be a nearly impossible task for any software. Moreover, numerical simulations that do take the non-commutativity 
\begin{equation}
[{\bf L}({\bf a}),\,{\bf L}({\bf b})]=-2\,{\bf L}({\bf a}\times{\bf b}) \label{locspin}
\end{equation}
of the spin variables ${\bf L(s)}$ into account are known to reproduce sinusoidal correlations violating Bell inequality, as reported in the references [8, 9, 10]. Consequently, Bell inequality is very likely to be violated in the experiment proposed above, just as it is violated in the microscopic quantum experiments [18].

\section{Concluding Remarks}

In this paper, we have compared Bell's local model for the singlet correlations set within ${\mathrm{I\!R}^3}$ with our local model set within $S^3$, and explained in detail why our model succeeds in reproducing the strong quantum correlations while Bell's model fails. For this purpose, we have focused on the singlet correlations involving spin angular momenta, whereas in most Bell-test experiments such as \cite{Aspect,Weihs-666} what is actually observed are correlations among the polarization states of entangled photons. It turns out, however, that the 3-sphere model we have discussed in this paper can easily accommodate such photon polarization states with appropriate changes in notation, as we have previously demonstrated in \cite{Failure} and in Chapter~8~of~\cite{Disproof}.

Now the important feature of the experiments performed in Orsay \cite{Aspect} and Innsbruck \cite{Weihs-666} using the polarization states of photons was that the measurement settings at the remote polarizers were changed {\it during} the flight of the two particles, thus removing any possibility of communication between the two ends of the experiment. The Innsbruck team precluded the possibility of communication between polarizers by using ultrafast random switching of the orientations of the polarizers. On each side of the experiment, a local computer registered the polarizer orientation and the result of each measurement, with timing monitored by an atomic clock, and the data was gathered and compared for
correlation only after the end of each run. These achievements prompted the principal investigator of the Orsay experiment to make the following comment \cite{Nature-666}:
\begin{quote}
I suggest we take the point of view of an external observer, who collects the data from the two distant stations at the end of the experiment, and compares the two series of results. This is what the Innsbruck team has done. Looking at the data a posteriori, they found that the correlation immediately changed as soon as one of the polarizers was switched, without any delay allowing for signal propagation ...
\end{quote}
But given the explanation we have offered in this paper by assuming the geometry of physical space to be that of a quaternionic 3-sphere, this immediate change in correlation should be no more puzzling than the sudden change in the color of Dr. Bertlmann's socks discussed by Bell \cite{Sock}, as we have explained elsewhere in considerable detail \cite{IEEE-2}. Consequently, in our view classical men like Hamilton, Grassmann, Clifford, and Hopf would not have been puzzled by the observed change in the correlation at all. They would have understood the results of Orsay \cite{Aspect} and Innsbruck \cite{Weihs-666} experiments in precisely the same local terms as we have explained them here; namely, in terms of the algebra of rotations in the physical space ${S^3}$, otherwise known as Clifford algebra \cite{Disproof,IJTP,RSOS,IEEE-1,IEEE-2,local,Failure}.

While in this paper we have focused only on the correlations predicted by the entangled singlet state \cite{IEEE-1}, elsewhere \cite{RSOS,local} we have developed a comprehensive local-realistic framework for understanding the origins of {\it all} quantum correlations in terms of the algebraic representation space $S^7$ of the quaternionic 3-sphere $S^3$. This demonstrates, constructively, that local hidden variable theories are not ruled out by the Bell-test experiments. Moreover, in \cite{Begs,Oversight} we have also explained why Bell's theorem fails for the non-commuting local spin variables that respect the relation (\ref{locspin}). While a systematic analysis of the reasons for the failure of Bell's theorem for non-commuting variables is beyond the scope of the present paper (because in this paper we are concerned with only a {\it single} expectation value (\ref{65a}) rather than a sum of expectation values such as (\ref{combi}) required to prove Bell's theorem), in \cite{Begs} we have shown that the traditional view of the strength $\pm2\sqrt{2}$ derived in (\ref{44final}) as due to non-local influences is meaningful only if one erroneously insists on linear additivity of expectation values of non-commuting observables for {\it individual} dispersion-free states. Once this error is corrected, the origin of the strength $\pm2\sqrt{2}$ of correlations can be traced to the Clifford-algebraic and geometric properties of physical space \cite{Begs,Oversight}.

\appendix

\section{Proofs of the Equalities (\ref{36-unm}) and (\ref{40-rtc})} \label{appa}

In this appendix we prove the equalities (\ref{36-unm}) and (\ref{40-rtc}), which amounts to proving that the
``product of limits equal to limits of product'' rule holds for quaternions and bivectors. To that end, we begin with the left-hand side of (\ref{36-unm}):
\begin{align}
\bigg[&\lim_{{\mathbf s}_1\,\rightarrow\,\mu_1{\mathbf a}}\left\{-\,{\mathbf D}({\mathbf a})\,{\mathbf L}({\mathbf s}_1)\right\}\bigg]\left[\lim_{{\mathbf s}_2\,\rightarrow\,\mu_2{\mathbf b}}\left\{+\,{\mathbf L}({\mathbf s}_2)\,{\mathbf D}({\mathbf b})\right\}\right] \notag \\
&=\bigg[\lim_{{\mathbf s}_1\,\rightarrow\,\mu_1{\mathbf a}}\left\{-I_3{\mathbf a}\,I_3{\mathbf s}_1\right\}\bigg]\left[\lim_{{\mathbf s}_2\,\rightarrow\,\mu_2{\mathbf b}}\left\{+I_3{\mathbf s}_2\,I_3{\mathbf b}\right\}\right] \\
&=\bigg[\lim_{{\mathbf s}_1\,\rightarrow\,\mu_1{\mathbf a}}\left\{-(I_3)^2\,{\mathbf a}\,{\mathbf s}_1\right\}\bigg]\left[\lim_{{\mathbf s}_2\,\rightarrow\,\mu_2{\mathbf b}}\left\{+(I_3)^2\,{\mathbf s}_2\,{\mathbf b}\right\}\right] \\
&=\bigg[\lim_{{\mathbf s}_1\,\rightarrow\,\mu_1{\mathbf a}}\left\{+\,{\mathbf a}\,{\mathbf s}_1\right\}\bigg]\left[\lim_{{\mathbf s}_2\,\rightarrow\,\mu_2{\mathbf b}}\left\{-\,{\mathbf s}_2\,{\mathbf b}\right\}\right] \\
&=\,\left[\mu_1{\mathbf a}\,{\mathbf a}\right]\left[-\mu_2{\mathbf b}\,{\mathbf b}\right] \\
&=-\mu_1\mu_2\,, \label{id1}
\end{align}
where we have used the fact that all vectors involved in the model are unit vectors and the fact that the pseudoscalar $I_3$ commutes with all other elements of $\mathrm{Cl}_{3,0}$ and squares to $-1$. Similarly, the right-hand side of (\ref{36-unm}) simplifies to
\begin{align}
\lim_{\substack{{\mathbf s}_1\,\rightarrow\,\mu_1{\mathbf a} \\ {\mathbf s}_2\,\rightarrow\,\mu_2{\mathbf b}}}&\left[\left\{-\,{\mathbf D}({\mathbf a})\,{\mathbf L}({\mathbf s}_1)\right\}\left\{+\,{\mathbf L}({\mathbf s}_2)\,{\mathbf D}({\mathbf b})\right\}\right] \notag \\
&=\lim_{\substack{{\mathbf s}_1\,\rightarrow\,\mu_1{\mathbf a} \\ {\mathbf s}_2\,\rightarrow\,\mu_2{\mathbf b}}}\left[   \left\{-I_3{\mathbf a}\,I_3{\mathbf s}_1\right\}\left\{+I_3{\mathbf s}_2\,I_3{\mathbf b})\right\}\right] \\
&=\lim_{\substack{{\mathbf s}_1\,\rightarrow\,\mu_1{\mathbf a} \\ {\mathbf s}_2\,\rightarrow\,\mu_2{\mathbf b}}}\left[   \left\{-(I_3)^2\,{\mathbf a}\,{\mathbf s}_1\right\}\left\{+(I_3)^2\,{\mathbf s}_2\,{\mathbf b})\right\}\right] \\
&=\lim_{\substack{{\mathbf s}_1\,\rightarrow\,\mu_1{\mathbf a} \\ {\mathbf s}_2\,\rightarrow\,\mu_2{\mathbf b}}}\left[   \left\{+\,{\mathbf a}\,{\mathbf s}_1\right\}\left\{-\,{\mathbf s}_2{\mathbf b})\right\}\right] \\
&=\lim_{\substack{{\mathbf s}_1\,\rightarrow\,\mu_1{\mathbf a} \\ {\mathbf s}_2\,\rightarrow\,\mu_2{\mathbf b}}}\left[-\,{\mathbf a}\,{\mathbf s}_1\,{\mathbf s}_2{\mathbf b}\right] \\
&=\left[-\mu_1{\mathbf a}\,{\mathbf a}\,\mu_2{\mathbf b}\,{\mathbf b}\right] \\
&=-\mu_1\mu_2. \label{id2}
\end{align}
Since the right-hand sides of (\ref{id1}) and (\ref{id2}) are equal, ``the product of limits equal to limits of product'' rule holds.

Analogously, we can also prove the equality (\ref{40-rtc}) by simplifying its left-hand side and right-hand side, as follows:
\begin{align}
&\left[\lim_{{\mathbf s}_2\,\rightarrow\,\mu_2{\mathbf b}}\left\{+\,{\mathbf L}({\mathbf s}_2)\,{\mathbf D}({\mathbf b})\right\}\right]\left[\lim_{{\mathbf s}_1\,\rightarrow\,\mu_1{\mathbf a}}\left\{-\,{\mathbf D}({\mathbf a})\,{\mathbf L}({\mathbf s}_1)\right\}\right] \notag \\
&=\bigg[\lim_{{\mathbf s}_2\,\rightarrow\,\mu_2{\mathbf b}}\left\{+I_3{\mathbf s}_2\,I_3{\mathbf b}\right\}\bigg]\left[\lim_{{\mathbf s}_1\,\rightarrow\,\mu_1{\mathbf a}}\left\{-I_3{\mathbf a}\,I_3{\mathbf s}_1\right\}\right] \\
&=\bigg[\lim_{{\mathbf s}_2\,\rightarrow\,\mu_2{\mathbf b}}\left\{+(I_3)^2\,{\mathbf s}_2\,{\mathbf b}\right\}\bigg]\left[\lim_{{\mathbf s}_1\,\rightarrow\,\mu_1{\mathbf a}}\left\{-(I_3)^2\,{\mathbf a}\,{\mathbf s}_1\right\}\right] \\
&=\bigg[\lim_{{\mathbf s}_2\,\rightarrow\,\mu_2{\mathbf b}}\left\{-\,{\mathbf s}_2\,{\mathbf b}\right\}\bigg]\left[\lim_{{\mathbf s}_1\,\rightarrow\,\mu_1{\mathbf a}}\left\{+\,{\mathbf a}\,{\mathbf s}_1\right\}\right] \\
&=\,\left[-\mu_2{\mathbf b}\,{\mathbf b}\right]\left[+\mu_1{\mathbf a}\,{\mathbf a}\right] \\
&=-\mu_1\mu_2 \label{id11}
\end{align}
and
\begin{align}
\lim_{\substack{{\mathbf s}_2\,\rightarrow\,\mu_2{\mathbf b} \\ {\mathbf s}_1\,\rightarrow\,\mu_1{\mathbf a}}}&\left[\left\{-{\mathbf D}({\mathbf b})\,{\mathbf D}({\mathbf a})\right\}\left\{\,{\mathbf L}({\mathbf s}_1)\,{\mathbf L}({\mathbf s}_2)\right\}\right] \notag \\
&=\lim_{\substack{{\mathbf s}_2\,\rightarrow\,\mu_2{\mathbf b} \\ {\mathbf s}_1\,\rightarrow\,\mu_1{\mathbf a}}}\left[   \left\{-I_3{\mathbf b}\,I_3{\mathbf a}\right\}\left\{I_3{\mathbf s}_1\,I_3{\mathbf s}_2)\right\}\right] \\
&=\lim_{\substack{{\mathbf s}_2\,\rightarrow\,\mu_2{\mathbf b} \\ {\mathbf s}_1\,\rightarrow\,\mu_1{\mathbf a}}}\left[   \left\{-(I_3)^2\,{\mathbf b}\,{\mathbf a}\right\}\left\{(I_3)^2\,{\mathbf s}_1\,{\mathbf s}_2)\right\}\right] \\
&=\lim_{\substack{{\mathbf s}_2\,\rightarrow\,\mu_2{\mathbf b} \\ {\mathbf s}_1\,\rightarrow\,\mu_1{\mathbf a}}}\left[   \left\{+\,{\mathbf b}\,{\mathbf a}\right\}\left\{-\,{\mathbf s}_1\,{\mathbf s}_2)\right\}\right] \\
&=\lim_{\substack{{\mathbf s}_2\,\rightarrow\,\mu_2{\mathbf b} \\ {\mathbf s}_1\,\rightarrow\,\mu_1{\mathbf a}}}\left[-\,{\mathbf b}\,{\mathbf a}\,{\mathbf s}_1\,{\mathbf s}_2\right] \\
&=\left[-\mu_1\mu_2\,{\mathbf b}\,{\mathbf a}\,{\mathbf a}\,{\mathbf b}\right] \\
&=\left[-\mu_1\mu_2\,{\mathbf b}\,{\mathbf b}\right] \\
&=-\mu_1\mu_2. \label{id22}
\end{align}
Since the right-hand sides of (\ref{id11}) and (\ref{id22}) are equal, ``the product of limits equal to limits of product'' rule holds.

\section{Questions and Answers} \label{QandA}

In no particular order, in this appendix I answer some questions concerning the $S^3$ model for the singlet correlations presented in this paper.

{\it Question 1}: While several good plausibility arguments are presented in the Introduction of the paper, they do not {\it prove} that physical space should be modeled as $S^3$ rather than $\mathrm{I\!R}^3$. How, then, is that assumption justified in the absence of proof? 

{\it Answer 1}: Apart from the plausibility arguments presented in the Introduction, there are several good reasons that justify the assumption of $S^3$. While this assumption is usually not made in the context of Bell’s theorem, the fact that quaternions play a fundamental role in understanding the algebra, geometry, and topology of physical
space is well-known since the works of Hamilton and Clifford in the 19th century and those of Pauli and  Dirac in the 20th
century. Dirac’s belt trick and Feynman’s plate trick are often used pedagogically to illustrate this fact. It has contributed to the formulation of Hypothesis~1 stated in the paper, which is a formal statement of my view that the best terrestrial evidence for $S^3$ as the geometry of physical space is the observed strong correlations in the Bell-test experiments.

But perhaps the strongest justification for the assumption of $S^3$ as physical space stems from the fact that the traditional interpretation of Bell's theorem is recovered from the $S^3$ model within the flat geometry of ${{\mathrm{I\!R}^3}}$ in which the characteristically non-trivial algebra, geometry, and topology of ${S^3}$ are absent. Consequently, the absolute upper bound of 2 on the Bell-CHSH combination (1) of expectation values is respected within $\mathrm{I\!R}^3$, as I have demonstrated, for example, in [9] and [14]. In fact, it is not difficult to demonstrate that the results presented in the current paper also reproduce the traditional interpretation of Bell's theorem in the flat geometry of ${{\mathrm{I\!R}}^3}$, which is usually taken for granted in the context of Bell's theorem. There are several different ways of appreciating this fact, as explained in Section~X of [9], each providing a different insight into how the standard interpretation of Bell's theorem is recovered in ${{\mathrm{I\!R}}^3}$.

As demonstrated in Section VI of [9], one way to appreciate it is by analyzing $S^3\rightarrow\mathrm{I\!R}^3$ limit in the event-by-event numerical simulations of the strong singlet correlations. A second way to appreciate it is by setting the parallelizing torsion ${{\mathscr T}}$ in $S^3$ to zero, as demonstrated in Section IX of [9], which reduces $S^3$ to $\mathrm{I\!R}^3$ as well as the absolute bound of $2\sqrt{2}$ on the Bell-CHSH sum (1) of expectation values to 2. A third way to appreciate it is by recognizing that the eigenvalue of the sum of quantum mechanical operators involved in Bell-CHSH inequalities and Bell-test experiments necessarily includes a purely geometrical contribution stemming from the $S^3$ geometry of physical space, and when that contribution is set to zero the absolute bound of $2\sqrt{2}$ on the Bell-CHSH sum (1) reduces to 2, as shown in [14].

But perhaps the best way to appreciate the recovery of the traditional interpretation of Bell's theorem within $\mathrm{I\!R}^3$ from the $S^3$ model is geometrical, by recognizing that the strong singlet correlations $-\cos(\eta_{{\mathbf a}{\mathbf b}})$ are a direct consequence of M\"obius-like twists in the Hopf bundle of $S^3$, as explained in Section~\ref{Hopf}. 

{\it Question 2}: Does the 3-sphere hypothesis proposed in the paper mean that an observationally flat universe (which is the view preferred by the presently dominant $\Lambda$-CDM model of cosmology) would essentially disqualify the proposed $S^3$ model?

{\it Answer 2}: No. The hypothesis proposed in the Introduction of the paper stands on its own. It has strong support from the cosmological observations, and from the fact that the 3-sphere is a permitted spatial part of one of the well-known solutions of Einstein’s field equations of general relativity, but it is not {\it dependent} on this evidence from the cosmological observations.

{\it Question 3}: The topology of the quaternionic 3-sphere is a global property of $S^3$. Does not that make the $S^3$ model a non-local realistic model rather than a local-realistic one as claimed?

{\it Answer 3}: The topology of $S^3$ is indeed a global property. But that does not make the $S^3$ model non-local realistic, at least for two fundamental reasons. To begin with, $S^3$, when viewed as physical space as in the model, is a spatial part of a well-known solution of Einstein's field equations of general relativity, which is a locally causal theory. Moreover, within the context of Bell's theorem, local causality is very specifically defined in terms of measurement results, such as the functions ${\mathscr A}({\mathbf a},{\mathbf s}^i_1)=\pm1$ and ${\mathscr B}({\mathbf b},{\mathbf s}^i_2)=\pm1$ defined in equations (17) and (23). If these functions are local, then the model is said to be local. And it is easy to verify that the functions are indeed local in the sense espoused by Einstein and later mathematically formulated by Bell [4]: Apart from the hidden variable ${\bf s}^i_1$, the result ${{\mathscr A}=\pm1}$ depends {\it only} on the measurement direction ${\bf a}$, chosen freely by Alice, regardless of Bob's actions. And, analogously, apart from the hidden variable ${\bf s}^i_2$, the result ${{\mathscr B}=\pm1}$ depends {\it only} on the measurement direction ${\bf b}$, chosen freely by Bob, regardless of Alice's actions. In particular, the function ${{\mathscr A}({\bf a},\,{\bf s}^i_1)}$ {\it does not} depend on ${\bf b}$ or ${\mathscr B}$ and the function ${{\mathscr B}({\bf b},\,{\bf s}^i_2)}$ {\it does not} depend on ${\bf a}$ or ${\mathscr A}$. Moreover, the hidden variables ${\bf s}^i_1$ and ${\bf s}^i_2$ do not depend on ${\bf a}$, ${\bf b}$, ${\mathscr A}$, or ${\mathscr B}$. Therefore, the model presented in the paper is manifestly local-realistic.

{\it Question 4}: Is it possible that $S^3$ geometry is relevant only for the space of entangled particles rather than physical space?

{\it Answer 4}: That would be a different hypothesis, not the one considered in the paper. It would amount to assuming that $S^3$ is an internal space, as in gauge theories. That hypothesis would not allow us to overcome Bell's theorem while respecting both the conditions of local realism and the observational constraints set by the Bell-test experiments. In other words, under that hypothesis, it would not be possible to derive the strong correlations (63), together with vanishing averages (32) and (33) for the separate results observed by Alice and Bob, while maintaining the strict locality condition explained above. This is because Bell's theorem remains valid within the flat geometry of $\mathrm{I\!R}^3$, as explained below and in Section~X of Ref.~[9].

{\it Question 5}: What is meant by the statement ``during the free evolution, the spins do not change in either their senses or directions''? If the spins do not change in their senses of directions, then in what sense is there ``evolution" of the spins?

{\it Answer 5}: Within the context of EPR-Bohm experiments, ``free evolution'' means the motion of the emerging spins free from interactions before they reach the detectors. Consequently, between the source and the detectors, the senses and directions of the spins are preserved. However, they are still propagating from the source to the detectors during this time. Soon after the constituent fermions emerge from the source as shown in Fig.~1, they cease to interact with each other appreciably but continue their journey toward the detectors. Moreover, since the spin system is assumed to be isolated, there is no external interaction between the spins and any other physical system until they interact with the two detectors. Also, their combined spin angular momentum would remain equal to zero, thanks to the law of the conservation of angular momentum. Therefore, the two spins do not change in their senses or directions while evolving between the source and the detectors. In other words, the EPR-Bohm correlations are purely kinematical effects. This is well understood since the first analysis of the singlet correlations using fermionic spins was carried out by Bohm in Section~22.16 of his 1951 book on quantum theory \cite{Bohm}. 

{\it Question 6}: The Stern-Gerlach device has a non-uniform electromagnetic field that aligns incoming charged particles to its orientation. In the $S^3$ model, the physics of this interaction and the nature of the particle itself are ignored, with the process of alignment being modeled with limit functions. But while it may be natural to mathematically represent spins of fermionic particles with bivectors, there is a vast difference between a Stern-Gerlach detector and a particle spin, and yet in the $S^3$ model both are represented with bivectors. Is that justified?

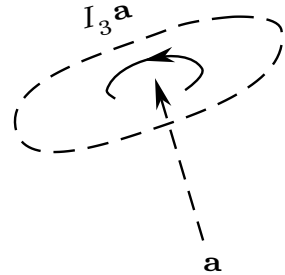
\begin{figure}[t]
\hrule
\scalebox{1.6}{
\begin{pspicture}(0.0,-2.1)(-6.4,-5.0)

\rotateleft{
\rotateleft{

\pscurve[linewidth=0.2mm,linestyle=dashed](3.25,2.64)(3.75,2.73)(5.05,3.6)(4.0,3.6)(2.85,2.75)(3.25,2.64)

\rput{180}(3.425,4.7){\scriptsize {${\bf a}$}}

\psline[linewidth=0.2mm,linestyle=dashed,arrowinset=0.3,arrowsize=2pt 3,arrowlength=2]{<-}(3.9,3.15)(3.5,4.5)

\pscurve[linewidth=0.2mm]{-}(3.65,3.25)(3.51,3.07)(4.3,3.25)(4.23,3.32)

\psline[linewidth=0.2mm,arrowinset=0.3,arrowsize=2pt 3,arrowlength=2]{->}(3.9,2.985)(4.0,3.005)

\rput{194}(4.3,2.65){\scriptsize {${I_{_3}{\bf a}}$}}

}
}

\end{pspicture}}
\hrule
\smallskip
\caption{A bivector, which represents a continuous binary rotation about some axis, is understood as an abstraction of a directed plane segment, with only a magnitude and a sense of rotation---{\it i.e.}, clockwise (${-}$) or counterclockwise (${+}$). Neither the depicted oval shape of its plane, nor its axis of rotation such as ${\bf a}$, is an intrinsic part of the bivector ${I_3{\bf a}}$. After~[9].}
\label{fig5}
\hrule
\end{figure}

{\it Answer 6}: Yes, that is perfectly justified and natural. The details of a specific experimental device are not relevant in the $S^3$ model, just as they are not relevant in Bell's local model set within $\mathrm{I\!R}^3$. In Bell's local model, what is represented by a vector, such as ``${\bf a}$'' appearing in the argument of the functions ${\mathscr A}({\bf a},{\bf s}_1^i)$, is an orientation of a detector, in three-dimensional physical space, mathematically modeled as $\mathrm{I\!R}^3$. The detectors such as the Stern-Gerlach apparatus themselves are, of course, complicated physical objects, whereas a vector is a geometrical object that models only the orientation of a detector in physical space. That is fine when physical space is a non-compact flat space such as $\mathrm{I\!R}^3$. The compact space $S^3$, however, has very different properties. I have defined it as a set of quaternions: 
\begin{equation}
S^3:=\left\{\,{\bf q}(\beta,\,{\mathbf r})=\varrho_r\exp\left[{\mathbf J}({\mathbf r})\,\frac{\beta}{2}\right]
\Bigg|\;\left|\left|\,{\bf q}\left(\beta,\,{\mathbf r}\right)\,\right|\right|=\varrho_r\right\}.
\end{equation}
As it is {\it constituted} by quaternions, an orientation of a detector within $S^3$ is best represented by a pure quaternion or bivector such as ${\bf D}({\bf a}):=I_3{\bf a}$. Recall that a bivector is also a geometrical object, much like a vector within $\mathrm{I\!R}^3$. It is an abstraction of an oriented plane segment with only three properties: magnitude, direction, and sense of rotation (see Fig.~\ref{fig5}). A bivector ${\bf D}({\bf a}):=I_3{\bf a}$ within $S^3$ is thus a counterpart of a vector ``${\bf a}$'' within $\mathrm{I\!R}^3$. As such, it provides the most adequate representation of an orientation of a detector in $S^3$. A vector cannot serve this purpose within $S^3$.

{\it Question 7}: What is the motivation for using mathematical limits to model the physical processes of detecting spin values?

{\it Answer 7}: Recall that the measurement function ${{\mathscr A}({\bf a},\,{\bf s}^i_1)}$ for Alice is defined in Bell's local model using the sign function: 
\begin{equation}
{\mathscr A}({\bf a},\,{\bf s}^i_1)=\text{sign}({\bf a}\cdot{\bf s}^i_1)=\pm1. \label{def}
\end{equation}
As noted before, the details of specific experimental devices, or the physical workings of the measurement instruments, are not relevant in formulating Bell’s local model, set within $\mathrm{I\!R}^3$. What is modeled in (\ref{def}) is the experimental fact that measurement instruments like any Stern–Gerlach device detect a spin by aligning its axis of rotation ${\bf s}^i_1$ with the vector $\bf{a}$ that specifies the orientation of the device in space. This produces an observable spot projected onto a screen, indicating that the spin is ``up'' or ``down'' along the direction $\bf{a}$, which is then mathematically represented by a number $+1$ or $-1$. Any other details of the physical workings of the device are not relevant to Bell's definition (\ref{def}) of the measurement results. But this definition is for the detection processes taking place within $\mathrm{I\!R}^3$. Its use of sign function is not valid for the detection processes taking place within $S^3$. Since the counterparts of the vectors $\bf{a}\in\mathrm{I\!R}^3$ and $\bf{s}_1\in\mathrm{I\!R}^3$ within $S^3$ are the bivectors ${\mathbf D}({\mathbf a})$ and ${\mathbf L}({\mathbf s}_1)$ representing rotations about the axes $\bf{a}$ and $\bf{s}_1$ (see Fig.~\ref{fig5}), what is required is a limit of the product quaternion ${\mathbf D}({\mathbf a})\,{\mathbf L}({\mathbf s}_1)$ representing the physical interaction between ${\mathbf D}({\mathbf a})$ and ${\mathbf L}({\mathbf s}_1)$ to produce the discrete results $\pm1$ observed by Alice. Fortunately, definition (\ref{def}) can be amended to accommodate this for the detection processes taking place within $S^3$, so that the correct definition of the measurement function ${\mathscr A}({\bf a},\,{\bf s}^i_1)$ is
\begin{align}
S^3\ni{\mathscr A}({\mathbf a},{\mathbf s}^i_1)\,&=\lim_{{\mathbf s}_1\,\rightarrow\,\mu_1{\mathbf a}}\left\{-\,{\mathbf D}({\mathbf a})\,{\mathbf L}({\mathbf s}_1)\right\} \label{797-nmn} \\
&=\lim_{{\mathbf s}_1\,\rightarrow\,\mu_1{\mathbf a}}\left\{-(I_3{\mathbf a})(I_3{\mathbf s}_1)\right\} \\
&=\mu_1\,, \label{sub}
\end{align}
where the quaternion $\left\{-\,{\mathbf D}({\mathbf a})\,{\mathbf L}({\mathbf s}_1)\right\}$ is an element of set $S^3$,
\begin{equation}
\mu_1 :=\text{sign}(\mathbf{a}\cdot\mathbf{s}^i_1)=\pm1,
\end{equation}
and the substitution rule for limits (and the fact that the unit bivector $I_3{\mathbf a}$ squares to $-1$) is used to derive the equality (\ref{sub}). Thus, definition (\ref{797-nmn}) reproduces Bell's definition (\ref{def}) in $\mathrm{I\!R}^3$, with the use of limit function accomplishing the experimentally required alignment of the vector $\bf{s}^i_1$ with the vector $\bf{a}$.  In other words, the use of limits to model the physical processes within $S^3$ of spin detections seen in Stern–Gerlach devices is not only required by the geometrical properties of $S^3$, but also natural. It is also worth noting that the substitution rule for limits used above, which essentially amounts to writing
\begin{equation}
\lim_{{\mathbf s}_1\,\rightarrow\,{\mathbf a}}\,I_3\,{\mathbf s}_1\,=\,I_3\,{\mathbf a}
\end{equation}
because ${\mathbf L}({\mathbf s}_1):=I_3\,{\mathbf s}_1$ and ${\mathbf L}({\mathbf a}):=I_3\,{\mathbf a}$, follows at once from the rules of Geometric Algebra. Moreover, it is easy to verify that the substitution rule for limits itself can be derived rigorously for any functions of vector quantities using the $\epsilon$--$\delta$ definition of limits from basic calculus \cite{Calculus}.

{\it Question 8}: Why are the directions of the constituent spins anti-aligned in Bell's model by setting ${\bf s}_1=-{\bf s}_2$, whereas in the $S^3$ model they are aligned to each other? There might be some subtlety in this regard, because if one of the observers is rotated by 180 degrees in order to face oncoming particles, then that will flip the horizontal directions of their coordinate axes. 

{\it Answer 8}: A fundamental difference between Bell's model set within $\mathrm{I\!R}^3$ and the quaternionic 3-sphere model set within $S^3$ is that, while in Bell's model the spin angular momenta are incorrectly represented by ordinary vectors such as ${\bf s}_1$ and ${\bf s}_2$ and the conservation of initial zero spin is stipulated as ${\bf s}_1+{\bf s}_2=0$ or as ${\bf s}_1=-{\bf s}_2$, in the $S^3$ model the spin angular momenta are correctly represented by the bivectors $-{\mathbf L}({\mathbf s}_1)$ and ${\mathbf L}({\mathbf s}_2)$ because they are rotating in opposite senses. As a result, the conservation of net zero spin is stipulated in equation (13) as
\begin{align}
-\,{\bf L}({\bf s}_1)\,+\,{\bf L}({\bf s}_2)=0 
\;&\Longleftrightarrow\;{\bf L}({\bf s}_1)={\bf L}({\bf s}_2) \notag \\
&\Longleftrightarrow\;{\bf s}_1=\,{\bf s}_2\,\equiv\,{\bf s}. \label{5555}
\end{align}
because $-{\mathbf L}({\mathbf s}_1)=-I_3{\bf s}_1$ and ${\mathbf L}({\mathbf s}_2)=I_3{\bf s}_2$, which is equivalent to the following conservation condition stated in equation (14):
\begin{equation}
{\bf L}({\bf s}_1)\,{\bf L}({\bf s}_2)=\,I_3{\bf s}\,I_3{\bf s}=(I_3)^2\,{\bf s}\,{\bf s}=-1.
\end{equation}
Once these mathematical relations are strictly followed, no verbal description with regard to what will happen in a given specific scenario such as ``what if one of the observers is rotated by 180 degrees to face the oncoming particles,'' {\it etc.} is necessary.

{\it Question 9}: Bell’s general argument and other impossibility arguments do not assume that space has the structure of $\mathrm{I\!R}^3$.

{\it Answer 9}: Although widely believed, this assertion is quite incorrect. Bell’s theorem and other impossibility arguments do implicitly assume that physical space has the structure of $\mathrm{I\!R}^3$. It is unfortunate that neither Bell nor other proponents of his theorem state this assumption explicitly. The only exception is a brief remark by Bell, quoted in the Introduction of the paper, which reads: “The space time structure has been taken as given here. How then about gravitation?” By ignoring the qualitative features of spacetime, Bell and his followers have neglected the geometrical and topological properties of physical space in their theorems, such as the non-commutativity of quaternions that constitute the 3-sphere. It is evident from the model considered in the paper that when these properties are taken into account, the observed strong singlet correlations can be easily derived.

{\it Question 10}: Bell’s general argument and other impossibility arguments do not rely on the particular local model discussed in Section~3 of Bell's 1964 paper [4] and in Peres's book [16].

{\it Answer 10}: This is not quite correct. The mathematical core of Bell’s theorem, while not dependent on the specific local model discussed in Section~3 of his paper, is nevertheless based on the correlations predicted by the singlet state. As such, the local model in Section~3 of his paper is of considerable interest. It is very much a part of the impossibility claim of the theorem.

\end{document}